\documentclass[aps,prb,twocolumn,showpacs,amsmath,superscriptaddress]{revtex4-1}
\usepackage{amssymb,amsmath}
\usepackage{mathrsfs}
\usepackage{graphicx}
\usepackage{float}
\usepackage{txfonts}
\usepackage[position=t,singlelinecheck=off,
caption=false]{subfig}
\usepackage[normalem]{ulem}
\usepackage{color}
\usepackage{bm}
\usepackage{xcolor}
\usepackage{pdfcomment}
\usepackage{mathrsfs}
\usepackage{graphicx}
\usepackage{dcolumn}
\usepackage{bm}
\usepackage{soul}
\setstcolor{red}

\newcommand{\ee}{\mathrm{e}}
\newcommand{\ii}{\mathrm{i}}

\begin{document}

\preprint{APS/123-QED}

\title{Photoinduced $\frac{1}{2}\frac{e^2}{h}$ and $\frac{3}{2}\frac{e^2}{h}$ conductance plateaus in topological insulator-superconductor heterostructures}

\author{Han Hoe \surname{Yap}}
\affiliation{NUS Graduate School for Integrative Sciences and Engineering, Singapore 117456, Republic of Singapore}
\affiliation{Department of Physics, National University of Singapore, Singapore 117551, Republic of Singapore}

\author{Longwen \surname{Zhou}}
\affiliation{Department of Physics, National University of Singapore, Singapore 117551, Republic of Singapore}

\author{Ching Hua \surname{Lee}}
\affiliation{Institute of High Performance Computing, Singapore 138632, Republic of Singapore}
\affiliation{Department of Physics, National University of Singapore, Singapore 117551, Republic of Singapore}

\author{Jiangbin \surname{Gong}} \email{phygj@nus.edu.sg}
\affiliation{Department of Physics, National University of Singapore, Singapore 117551, Republic of Singapore}
\affiliation{NUS Graduate School for Integrative Sciences and Engineering, Singapore 117456, Republic of Singapore}

\date{\today}

\begin{abstract}
{The past few years have witnessed increased attention to the quest for} {Majorana-like excitations in the condensed matter community.  As a promising candidate in this race, the one-dimensional chiral Majorana edge mode (CMEM) in   {topological insulator-superconductor heterostructures has gathered renewed interests} during recent months after an experimental breakthrough.   {In this paper, we study the quantum transport of topological insulator-superconductor hybrid devices subject to light-matter interaction  or general time-periodic modulation}. We report half-integer quantized conductance plateaus at $\frac{1}{2}\frac{e^2}{h}$ and $\frac{3}{2}\frac{e^2}{h}$ upon applying the so-called sum rule in the theory of quantum transport in Floquet topological matter. In particular, in a photoinduced topological superconductor sandwiched between two Floquet Chern insulators, it is found that for each Floquet sideband, the CMEM admits equal probability for normal transmission and local Andreev reflection over a wide range of parameter regimes, yielding half-integer quantized plateaus that resist static and time-periodic disorder.  The $\frac{3}{2}\frac{e^2}{h}$ plateau has not yet been computationally or experimentally observed in any other {superconducting system}, and indicates the {possibility to simultaneously create and manipulate multiple pairs of CMEMs by light}. The robust and half-quantized conductance plateaus, due to CMEMs at {quasienergies zero or half the driving frequency}, are both fascinating and subtle because they only emerge after a summation over contributions from all Floquet sidebands. {Such a distinctive transport signature} can thus serve as a hallmark of photoinduced CMEMs in topological insulator-superconductor junctions.}
\end{abstract}

\maketitle

\section{Introduction}
The search for excitations with Majorana-like properties in solid state systems is an active ongoing research topic\cite{Beenakker2,Alicea,Flensberg,Lutchyn,MajoranaReview}. A vast majority of efforts focuses on zero-dimensional Majorana bound states (MBS) at the ends of proximitized semiconductor nanowires\cite{DasSarma3,DasSarma1,DasSarma2,DasSarma4,Kouwenhoven,Kouwenhoven2,ZBP,FlensbergExperiment} or magnetic atom chains on superconductor substrates\cite{Yazdani-1,Yazdani0,Yazdani,Yazdani2,Yazdani3,MajoranaChain}. Another candidate in the race is the one-dimensional chiral Majorana edge modes (CMEM) on the surface of topological insulators\cite{Kane,SCZhang1} (TI). Recently, transport measurements\cite{SCZhangExperiment} have been carried out in quantum anomalous Hall insulators (QAHI) proximitized by s-wave superconductor. Half-quantized conductance plateaus were identified and interpreted as the signature of the existence of a pair of CMEM, a conclusion which has sparked vivid debates\cite{XGWen,JaySau,SCZhangReply} and inspired {  a number of stimulating} follow-up studies\cite{KTLaw,CKChiu,CKChiu2,XCXie,KTLaw2,MacDonald,QingFeng}.

Such intense attention is not unwarranted for: Majorana zero modes hold promising prospects for topological quantum computation\cite{SarmaQuantumComputation,StanescuBook} (TQC), and their discovery should demonstrate the ability of table-top experiments in simulating fundamental particles not yet found in nature\cite{Wilczek}. With such far-reaching impacts, the pursuit of Majorana fermions naturally spans beyond the realm of solid state materials. Be it on cold atom systems\cite{Zoller1,Zoller2,Zoller3} or photonic lattices\cite{Synthetic,Synthetic2}, proposals towards the realizations of Majorana fermions are regularly put forward.

A recurring theme among these approaches is the manipulation of electromagnetic waves\cite{Goldman}. Indeed, the revelation that light can interact with matter and give rise to nontrivial Floquet-Bloch states\cite{Lindner1} has propelled our understanding of topological phases\cite{Derek,Lindner2,Kundu2} and opened up a new avenue in the conception of novel topological matter\cite{OnDemand}. From the surge of research activities emerges the field of Floquet topological phases\cite{Kitagawa,Asboth,Zhao}, and rightfully so: Not only do they exhibit unusual bulk-edge correspondence\cite{Nathan,Titum}, the tunability of periodic driving fields also allows the generation of intriguing phases with large topological invariants\cite{JBGong1,Derek2,LWZhou2,TSXiong,Martin,OkaWeyl}. Of course, ingenious engineering finds its way also in non-photonic settings. Concepts of Floquet topological phases have since been extended to mechanical lattices\cite{Fleury,CHLee} and proposed in electrical circuits\cite{CHLee2}.

In the context of superconductors, Floquet systems have an additional bonus: Because of the torus topology of the Floquet-Brillouin zone, there is no distinction between quasienergies $\epsilon=\pm\hbar\Omega/2$ ({  $\Omega$ being the frequency of a periodic drive)}.
Thus zero energy no longer reigns supreme, because excitations $\gamma$ obeying the Majorana condition $(\gamma^\dag=\gamma)$\cite{Chamon,BenaReview,Beenakker,Elliott} can now be sought at {quasienergy} $\epsilon=\pm\hbar\Omega/2$ as well. In the literature, most studies concentrated on dissecting the topological features\cite{JBGong1,Raditya2,FloquetMajorana1,FloquetMajorana2,FloquetMajorana3,FloquetMajorana4}, while the transport properties of driven one-dimensional systems hosting MBS have also been investigated\cite{Kundu,MajoranaTransport2,FloquetMajoranaTransport}. However, studies on the quantum transport { of two-dimensional Floquet topological superconductors \cite{Sacramento} are still scarce,  not to mention the possible quantization signature of CMEMs}.

{  Motivated by the importance of active manipulation of CMEMs, as well as the fascinating interplay between light-matter interaction and topology,  we investigate the possibility of creating Floquet
CMEMs in hybrid topological devices, which consist of a Floquet topological superconductor (FTSC) sandwiched between two Floquet Chern insulators (FlCI). We focus on the transport signature by computationally studying the longitudinal DC conductance. We discover conductance plateaus over wide ranges and diverse choices of parameters at $\frac{1}{2}\frac{e^2}{h}$ and $\frac{3}{2}\frac{e^2}{h}$ upon invoking the so-called Floquet sum rule\cite{Kundu}, for CMEMs at quasienergy zero or $\pm\hbar\Omega/2$.  It is found that if there is a conductance plateau, then for all Floquet sidebands, the normal and Andreev processes always share equal probabilities.  Therefore, it can be asserted that the property of CMEMs being equal superpositions of electrons and holes\cite{SCZhang4} extends to all Floquet sidebands.    {Consequently, in a FlCI-FTSC junction, for each individual Floquet sideband}, the incoming chiral fermion from FlCI splits into two CMEMs, one propagating along the FTSC-vacuum interface, and the other along the FlCI-FTSC domain wall.}

{  The half-quantized conductance plateaus discovered in this work are found to be robust against static and time-periodic disorder. They are rather subtle because they emerge only after summing over contributions from all Floquet sidebands.  They can thus serve as a hallmark of photoinduced CMEMs in topological insulator-superconductor junctions. In particular, the conductance plateau at $\frac{3}{2}\frac{e^2}{h}$ marks the existence of more than a single pair of CMEMs, indicating a possible way to simultaneously create and manipulate multiple pairs of CMEMs by light. We note that the $\frac{3}{2}\frac{e^2}{h}$ conductance plateaus have not yet been computationally or experimentally observed in any other   {superconducting system}. In passing, we also elaborate on a few other aspects of
transport in Floquet topological junctions.  For example, we show that in general the Chern numbers of FTSC and FlCI in a hybrid junction cannot be used to predict whether CMEMs manifest as half-quantized conductance plateaus.  These results offer more insights into the pursuit of Majorana fermions and contribute to the general understanding of transport in Floquet hybrid devices}.

This paper is structured as follows. In Sec.~\ref{sec:Model}, {  we propose system Hamiltonians and driving protocols modeling certain time-periodic modulation such as light-matter interaction. Various junctions consisting of FlCI and FTSC are thus constructed}. In Sec.~\ref{sec:Method}, {  we briefly describe technical details regarding how DC conductance is calculated throughout this work. Section~\ref{sec:Results} presents our main findings on conductance plateaus with detailed supporting results, including the effects of disorder. In Sec.~\ref{sec:Discussions}, we elaborate on several aspects of quantum transport in Floquet topological hybrid devices. Section~\ref{sec:Summary} summarizes this work.}

\section{Model}\label{sec:Model}
We study the two-terminal quantum transport of hybrid structures consisting of a superconductor sandwiched between two non-superconducting matter attached to two metallic leads.
\begin{figure}
\includegraphics[width=.4\textwidth]{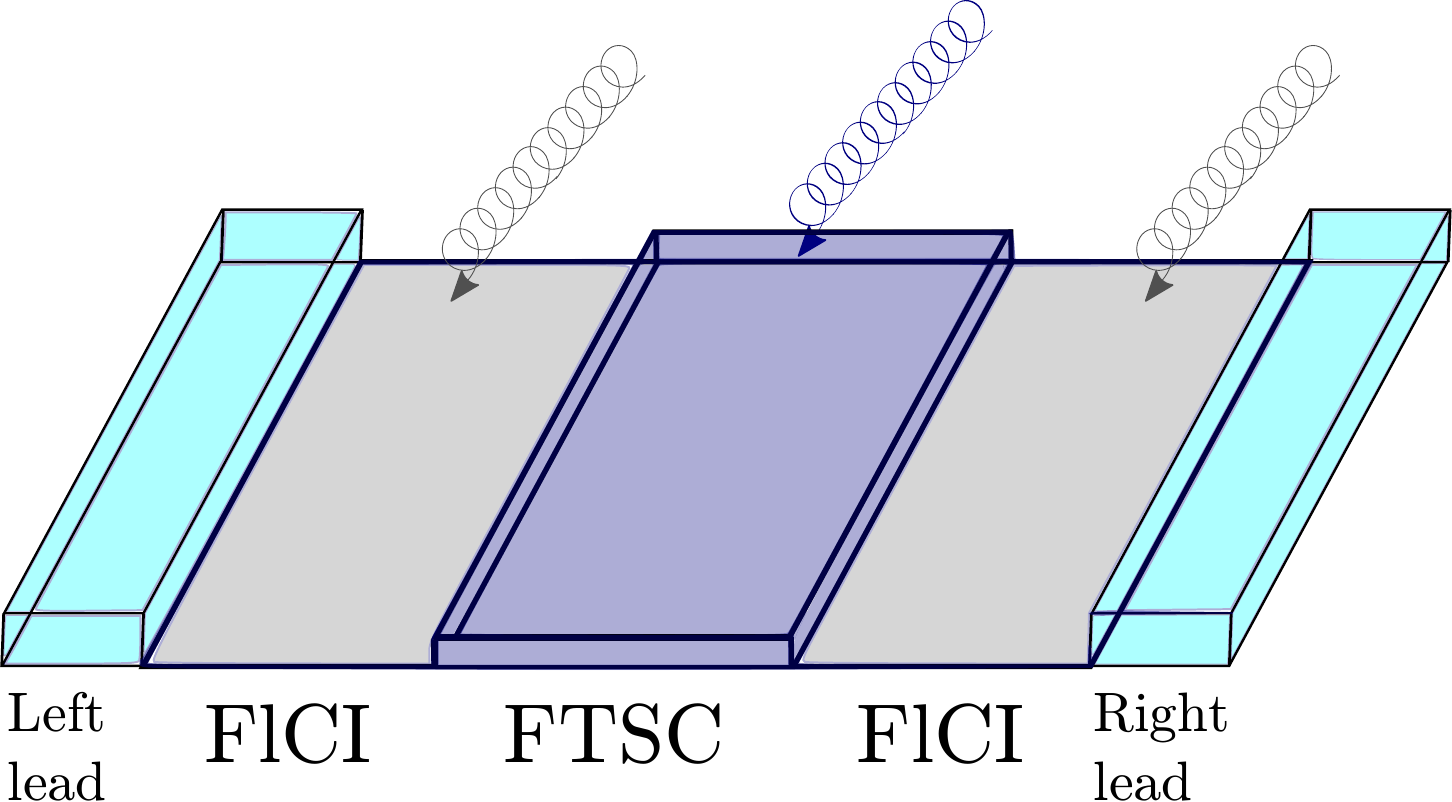}
\caption{Schematic of the model studied in this paper. The sample is composed of a Floquet topological superconductor, sandwiched between two Floquet Chern insulators, which are attached to two leads.}
\end{figure}
The entire sample is periodically driven at the same fundamental frequency $\Omega$, such that Floquet theory applies. The driving fields may or may not be homogeneous throughout the sample. The central superconducting region is described by a Hamiltonian $h_C$, proximitized by an s-wave superconductor of pairing strength $\Delta$. Therefore, within the Bogoliubov-de Gennes (BdG) formalism\cite{Sato}, the left/ right and central Hamiltonians read:
\begin{equation}
 H_{L/R}(\bm{k}) = \begin{bmatrix}
 h_{L/R}(\bm{k}) &  \\  & -h_{L/R}^*(-\bm{k})
 \end{bmatrix},
H_C(\bm{k}) = \begin{bmatrix}
h_C(\bm{k}) & -\ii \Delta \sigma_y \\ \ii \Delta \sigma_y & -h_C^*(-\bm{k})
\end{bmatrix}, \label{eq:Hamiltonian}
\end{equation}
where $\sigma_{x,y,z}$ are the Pauli matrices in sublattice or spin space.

\subsection{Static Hamiltonians}
To illustrate the general idea, we choose either the Haldane\cite{Haldane} or the Qi-Wu-Zhang (QWZ) model\cite{QiWuZhang} as the base systems, on which we add superconductivity and periodic drives. We write the Hamiltonians in momentum space, but in actual calculations we always work with open boundary conditions.
\subsubsection{Haldane model}
The Haldane model describes electrons on a honeycomb lattice with broken time reversal symmetry. As the first proposal for quantum anomalous Hall effect, it was recently realized by ultracold fermions in an optical lattice\cite{HaldaneExperiment}. Its Hamiltonian reads $h_C(\bm{k})=\bm{f}(\bm{k})\cdot\bm{\sigma}$, where:
\begin{equation}
\begin{split}
f_x(\bm{k}) &= t_1[1+\cos(\bm{k}\cdot\bm{a}_1)+\cos(\bm{k}\cdot\bm{a}_2)], \\
f_y(\bm{k}) &= t_1[\sin(\bm{k}\cdot\bm{a}_1)+\sin(\bm{k}\cdot\bm{a}_2)], \\
f_z(\bm{k}) &= 2t_2 \sin\Phi\big\{\sin(\bm{k}\cdot\bm{a}_1)-\sin(\bm{k}\cdot\bm{a}_2)-\sin[\bm{k}\cdot(\bm{a}_1-\bm{a}_2)] \big\}. \label{eq:graphene}
\end{split}
\end{equation}
Here, $t_{1,2}$ are the first and second neighbor hopping parameters, $\Phi$ is a phase factor, $\bm{a}_1=\sqrt{3}\hat{\bm{x}},\bm{a}_2=(\sqrt{3}\hat{\bm{x}}+3\hat{\bm{y}})/2$ are the primitive vectors of honeycomb lattice with zigzag edge, a geometry chosen for convenience.
\subsubsection{Qi-Wu-Zhang model}
The QWZ model is a two-band model that captures the essence of Chern insulators (CI). Physically, it describes half of the Bernevig-Hughes-Zhang model for quantum spin Hall effect in CdTe/HgTe heterostructure\cite{Bernevig}. With this choice, the static Hamiltonians read:
\begin{equation}
h_C(\bm{k}) = A\sin k_1 \sigma_x + A\sin k_2 \sigma_y + \big[m+4B(\cos k_1+\cos k_2)\big]\sigma_z,
\end{equation}
where $m,A,B$ are tunable material parameters.
\subsection{Periodic drives}
The effect of driving fields amounts to adding a periodic time dependence on the above Hamiltonians. In this paper, we discuss two types of protocols: continuous drive and periodic quenches.
\subsubsection{Continuous driving field}
For continuously driven models, the Hamiltonians are modified by Peierls' substitution, $\bm{k}\mapsto\bm{k}+e\bm{A}(t)$, where $-e<0$ is the electron charge, $\bm{A}(t)$ is the vector potential. On honeycomb lattices, we take the driving fields to be of the form $\bm{A}(t)=A\big[\sin(\Omega t)\bm{a}_1+\sin(\Omega t+\phi)\bm{a}_2\big]$, where $\phi$ is an angle characterizing the rotating electromagnetic waves.
\subsubsection{Periodic quenches}
A quench is an abrupt transition of a physical system. It is widely employed in experiments to {study thermalization, localization, and nonlinear dynamics\cite{Derek,Derek2,JBGong1,TSXiong,Quench,Quench2,Quench3,Quench4}}. When two quenches are periodically alternated, one obtains a two-cycle quench model, described by:
\begin{equation}
H(t) = \begin{cases}
h^{(1)},& 0<t<T_1\phantom{+2T_2}\mod{\mathcal{T}}, \\
h^{(2)}, & T_1<t<T_1+T_2\mod{\mathcal{T}},
\end{cases}
\end{equation}
where $\mathcal{T}=T_1+T_2=2\pi/\Omega$ is the period. Physically, periodic quenches of period $T_1+T_2$ are driving fields containing many different frequencies, commensurate with $\Omega=2\pi/(T_1+T_2)$. This is evidenced by the Fourier series: $H(t)=\sum_{k\in\mathbb{Z}}\exp(\ii k\Omega t)h_k$, where
\begin{equation}
h_k=\begin{cases}
  \dfrac{h^{(1)}T_1+h^{(2)}T_2}{T_1+T_2}, & k=0,
\\
\dfrac{1}{k\pi}\left[h^{(1)}\ee^{-\ii \frac{\Omega T_1}{2}}\sin\frac{k\Omega T_1}{2}
+h^{(2)}\ee^{-\ii \frac{\Omega(T_1+T_2)}{2}}\sin \frac{k\Omega T_2}{2}\right], & k\neq 0.
\end{cases}\label{eq:quench}
\end{equation}
Thus hypothetically one may envisage to implement approximate quenches using lights at frequencies commensurate with each other.

In the following, we work with units such that $e,\hbar$ are unity.

\section{Method}\label{sec:Method}
We apply the method of recursive Floquet-Green's function\cite{HHYap}, combined with the Floquet sum rule\cite{Kundu,Kundu2,Farrell,Farrell2} for the calculations of DC conductance. For undriven samples where superconductivity is present, one can show that if the left (right) lead is set at bias $V_{\mathrm{L/R}}=\pm V/2$, then the electrical current flowing in the sample reads\cite{mastersthesis,XCXie,XCXie2,XCXie3}:
\begin{equation}
\begin{split}
I&=\frac{e}{h}\int_{-\infty}^{\infty}\mathrm{d}E\big[g(E)+g^\mathrm{LAR}(E)\big]\big[f_{-V}(E)-f_V(E)\big],\label{eq:current}
\end{split}
\end{equation}
with the normal and local Andreev reflection coefficients given by:
\begin{equation}
\begin{split}
g(E) &= \mathrm{Tr}\left[G\Gamma^\mathrm{e}_\mathrm{R}G^\dag\Gamma^\mathrm{h}_\mathrm{L}\right],\\
g^\mathrm{LAR}(E) &= \mathrm{Tr}\left[G\Gamma^\mathrm{e}_\mathrm{R}G^\dag\Gamma^\mathrm{h}_\mathrm{R}\right],
\end{split}
\end{equation}
where $G=(E+ \ii\eta - H_\mathrm{BdG}-\Sigma_\mathrm{L}-\Sigma_\mathrm{R})^{-1}$ is the retarded Green's function, $\Sigma_{\mathrm{L}/\mathrm{R}}$ is the left/right self energy due to coupling with metallic leads, $\Gamma_\mathrm{L/R}^{\mathrm{e}/\mathrm{h}}=\ii P^{\mathrm{e}/\mathrm{h}}(\Sigma_\mathrm{L/R}-\Sigma^\dag_\mathrm{L/R})$ is the left/ right line-width function in electron/hole, $P^{\mathrm{e}/\mathrm{h}}=(1\pm\tau_z)/2$ is the electron/hole projector, $\tau_z$ is the Pauli matrix acting on Nambu space, and $f_V(E)=\{\exp[\beta(E+\mathrm{e}V)]+1\}^{-1}$ is the Fermi function at inverse temperature $\beta$. One can extract from Eq.~\eqref{eq:current} a differential conductance by linearizing the Fermi function, assuming a small bias $V$. For negligible temperature, at Fermi energy $E_\mathrm{F}$, one obtains the longitudinal conductance as:
\begin{equation}
g(E_\mathrm{F})=\frac{e^2}{h}\big[g(E_\mathrm{F})+g^\mathrm{LAR}(E_\mathrm{F})	\big].\label{eq:differentialConductance}
\end{equation}
For periodically driven systems, the time-averaged DC conductance for normal transmission reads\cite{Kohler}:
\begin{equation}
g_\mathrm{DC}(E)=\sum_{k\in\mathbb{Z}}\mathrm{Tr}\big[\bm{G}_{k0}(\bm{\Gamma}^{\mathrm{e}}_\mathrm{R})_{00}\bm{G}^\dag_{0k}(\bm{\Gamma}^{\mathrm{e}}_\mathrm{L})_{kk}\big],
\end{equation}
and analogously for the local Andreev reflection, where all Green's functions now are Floquet-Green's functions\cite{Oka}, satisfying, e.g.:
\begin{equation}
[(E+m\Omega)\bm{\delta}_{mk}-(\bm{H}_\mathrm{BdG})_{mk}-(\bm{\Sigma}_\mathrm{L}+\bm{\Sigma}_\mathrm{R})_{mk}]\bm{G}_{kn}(E) = \delta_{mn},
\end{equation}
and the subscripts mean: $(\bm{H}_\mathrm{BdG})_{mn}=\frac{\Omega}{2\pi}\int_0^{2\pi/\Omega}\mathrm{d}t \exp[\ii(m-n)\Omega t]H_\mathrm{BdG}(t)$. Finally, to faithfully reflect the topology of Floquet systems via transport calculations, a summation over different Floquet sidebands is performed,   {so that scattering channels at integer multiples of driving frequency away from the Fermi energy of the incoming state may be taken into account, as dictated by} the Floquet sum rule\cite{Kundu,Kundu2,Farrell,Farrell2}:
\begin{equation}
\bar{g}(E_\mathrm{F})= \sum_{n\in\mathbb{Z}} g_n(E_\mathrm{F}),
\end{equation}
where $g_n(E_\mathrm{F})=g(E_\mathrm{F}+n\hbar\Omega)$, valid for both normal and Andreev processes.

\section{Results}\label{sec:Results}
\subsection{Half-quantized plateau from two devices with mismatched Chern numbers}\label{sec:halfQuenched}
We begin with the s-wave superconducting Qi-Wu-Zhang (QWZ) model\cite{QiWuZhang}, which was first studied in Refs.~\cite{SCZhang1,SCZhang4}. For a single CMEM to manifest as half-quantized conductance, an important topological prerequisite is that one must have Chern numbers $\mathcal{C}=1$ and $\mathcal{N}=1$ for the two layers of Chern insulators and the topological superconductor in the middle respectively. This is because for $\mathcal{N}=2$, the incoming chiral ordinary fermion travels unimpeded\cite{SCZhang4}, whereas if $\mathcal{N}=0$, then it gets completely reflected\cite{SCZhang5}.

\begin{figure*}
\centering
\subfloat[]{
\includegraphics[width=.4\textwidth]{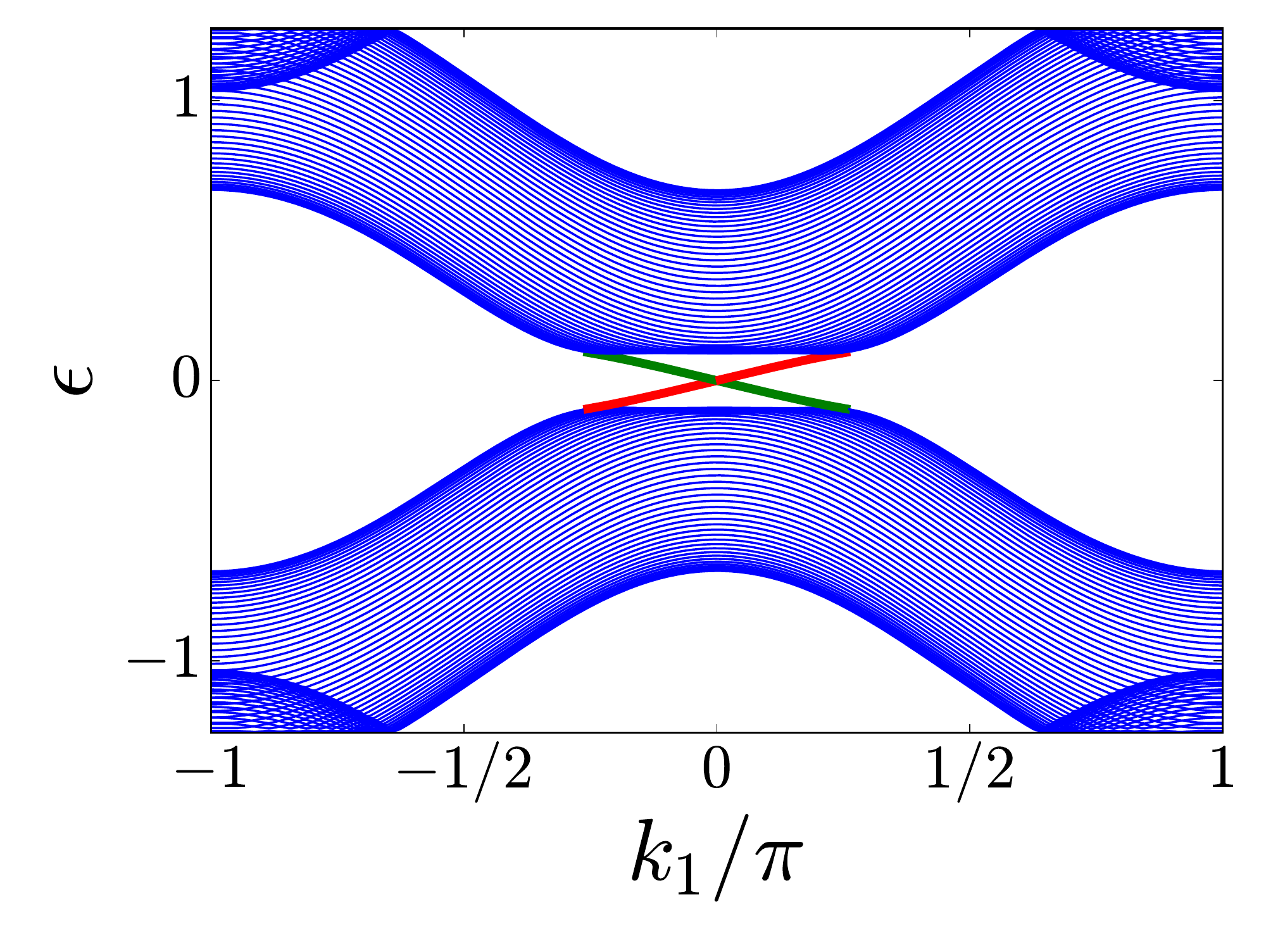}
}
\subfloat[]{
\includegraphics[width=.4\textwidth]{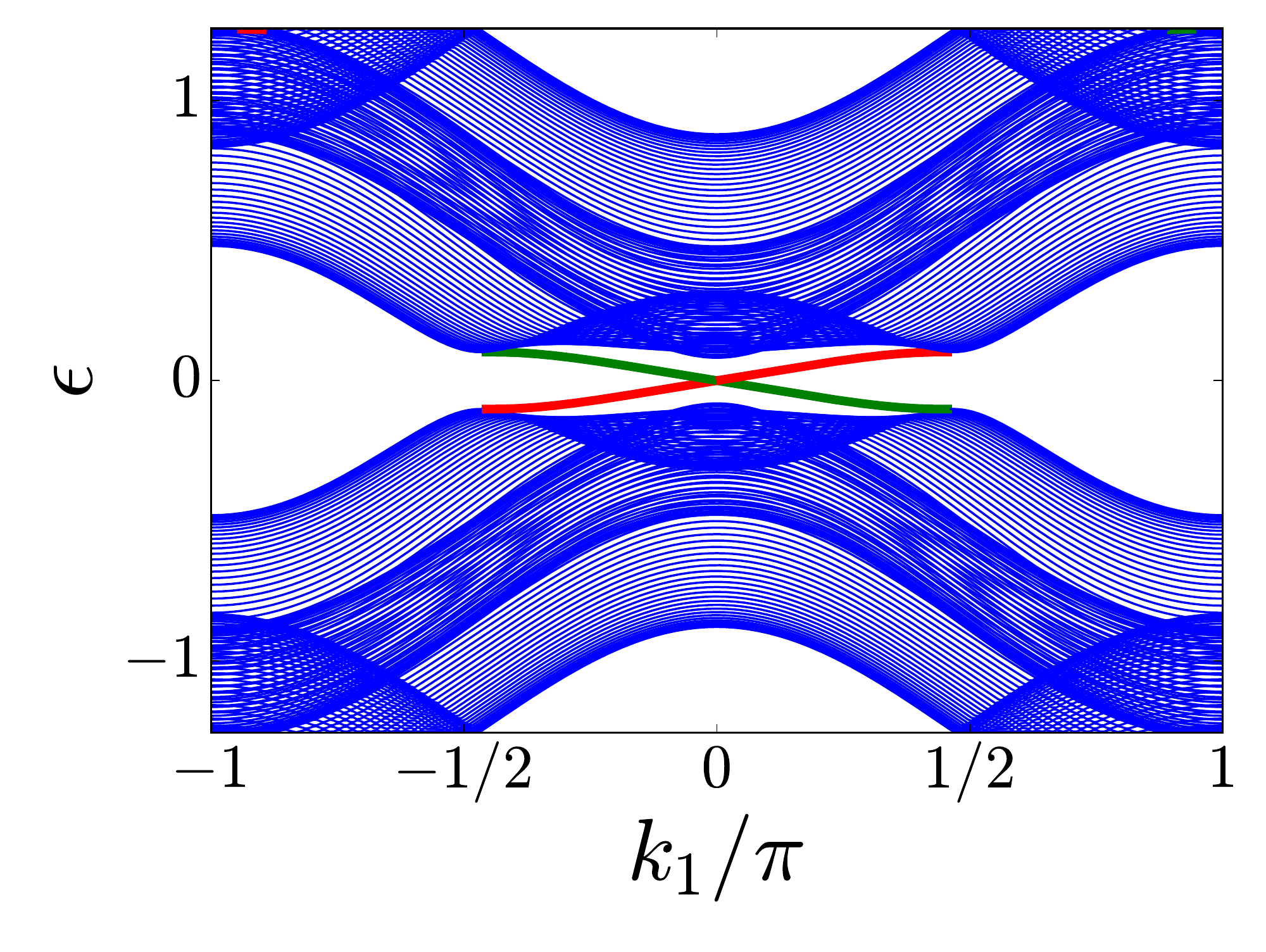}
}
\\
\centering
\subfloat[]{
\includegraphics[width=.4\textwidth]{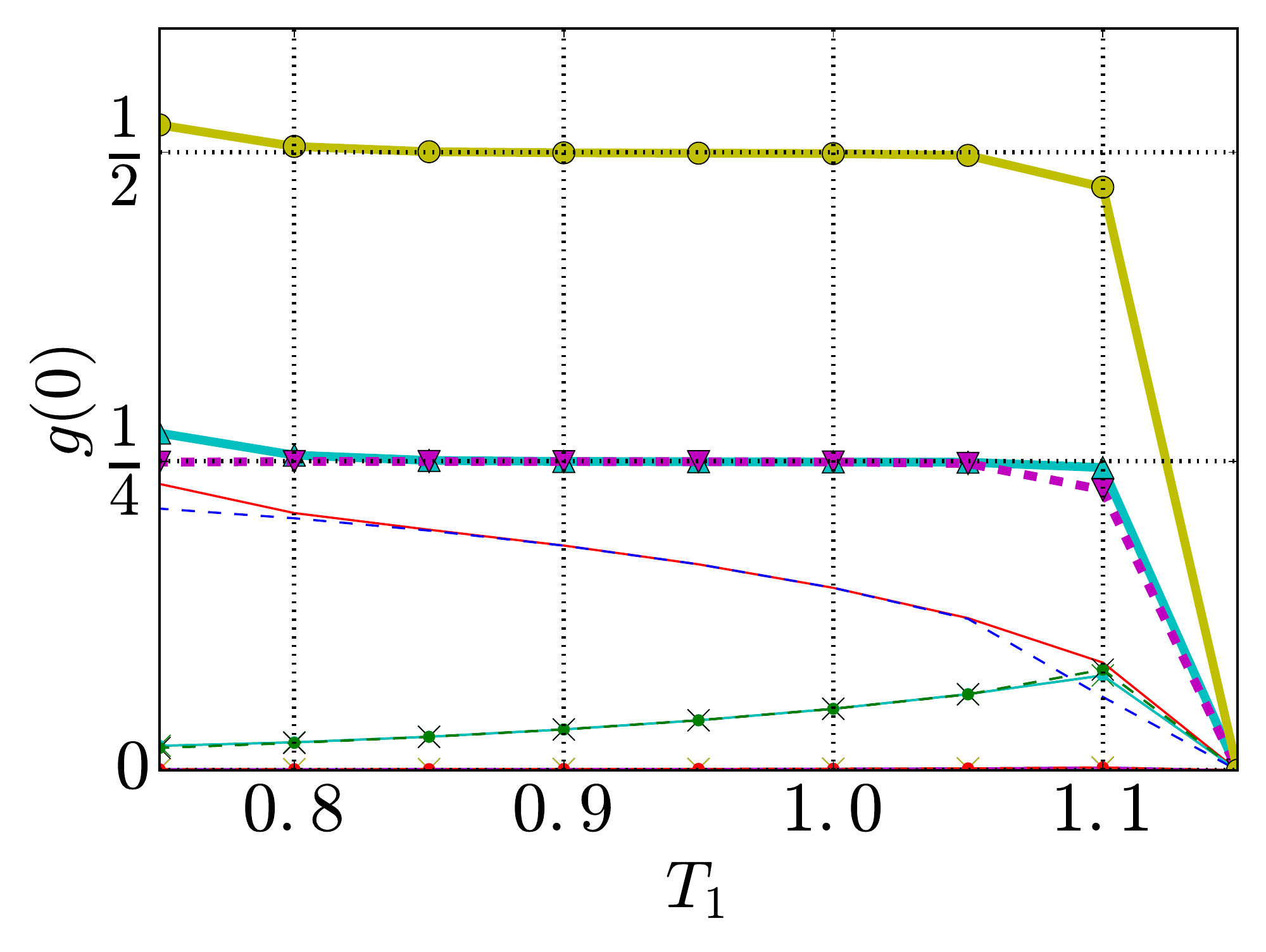}
}
\subfloat[]{
\includegraphics[width=.4\textwidth]{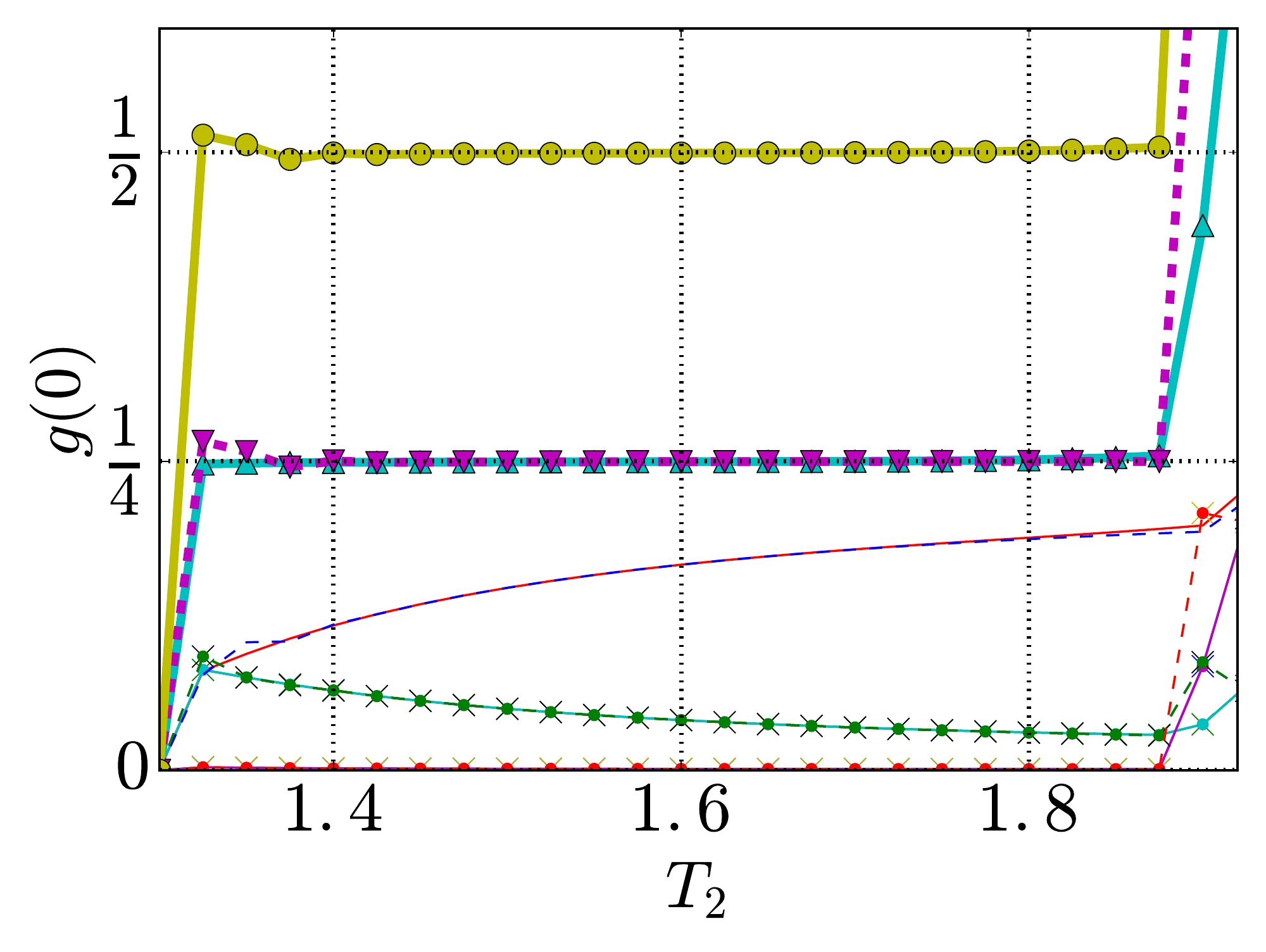}
}
\\
\centering
\subfloat[]{
\includegraphics[width=.7\textwidth]{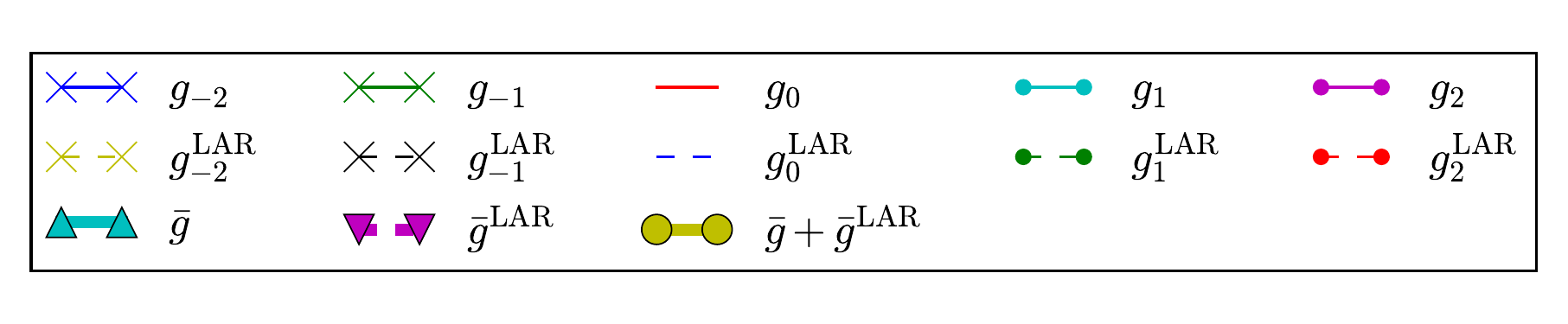}
}
\caption{Top panel: Spectra of (a): FlCI, (b): FTSC. Middle panel: DC conductance at $E_\mathrm{F}=0$ as the quench durations (c) $T_1$ and (d) $T_2$ vary. Bottom panel (e): Legends of plots for (c) and (d). Parameters of the two quenches: $m=(-2.1,1.6),A=(1,-1),B=(1,-1),\Delta=(1.6,-1.4),T_1=1,T_2=1.5$, applicable to both (a) and (b) except the pairing $\Delta$ which is present only at the FTSC (b).}
\label{fig:QXL}
\end{figure*}

Suppose now we periodically quench two devices with mismatched Chern numbers: $\mathcal{C}=1,\mathcal{N}=2$, and $\mathcal{C}=-1,\mathcal{N}=-2$. With two additional parameters, i.e. the quench durations $T_{1,2}$, the resultant device can explore a much larger topological phase diagram. In particular, with a judicious choice of $T_{1,2}$, the Floquet junction realizes a single chiral fermion at the FlCIs [Fig.~\ref{fig:QXL}(a)] and a single CMEM at the FTSC [Fig.~\ref{fig:QXL}(b)].

Hence, we perform DC conductance calculations at $E_\mathrm{F}=0$ by tuning the quench durations $T_1$ [Fig.~\ref{fig:QXL}(c)] and $T_2$ [Fig.~\ref{fig:QXL}(d)]. We observe that the individual sidebands for both normal transmission and local Andreev reflection $(g_n,g^\mathrm{LAR}_n)$ vary, but their sums $(\bar{g},\bar{g}^\mathrm{LAR})$ form plateaus of $1/4$, resulting in DC longitudinal plateaus of $1/2$.


\subsection{Photoinduced CMEM in graphene hybrid structure}\label{sec:graphene}
In the previous example, the two pre-quench junctions can actually realize a single pair of CMEM\cite{SCZhang1} with other choices of parameters. A natural question to ask is thus: In a device where the existence of CMEM is forbidden by symmetry, can periodic driving fields lift this restriction? Inspired by Ref.\cite{Kitagawa}, here we answer this question affirmatively by considering the dynamical generation of a single Floquet CMEM from a normal-superconducting graphene hybrid structure.

\begin{figure*}
\centering
\subfloat[]{
\includegraphics[width=.4\textwidth]{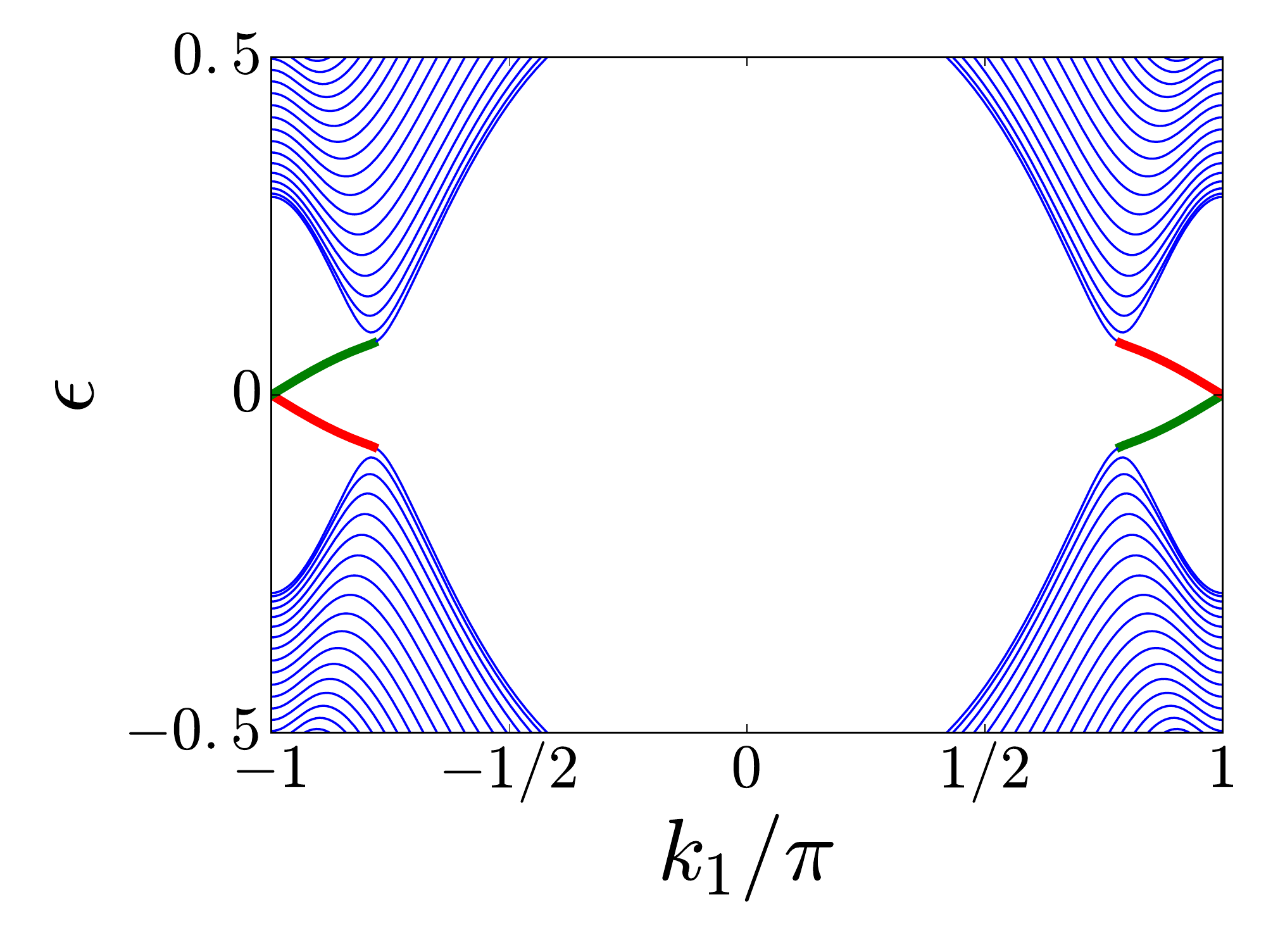}
}
\subfloat[]{
\includegraphics[width=.4\textwidth]{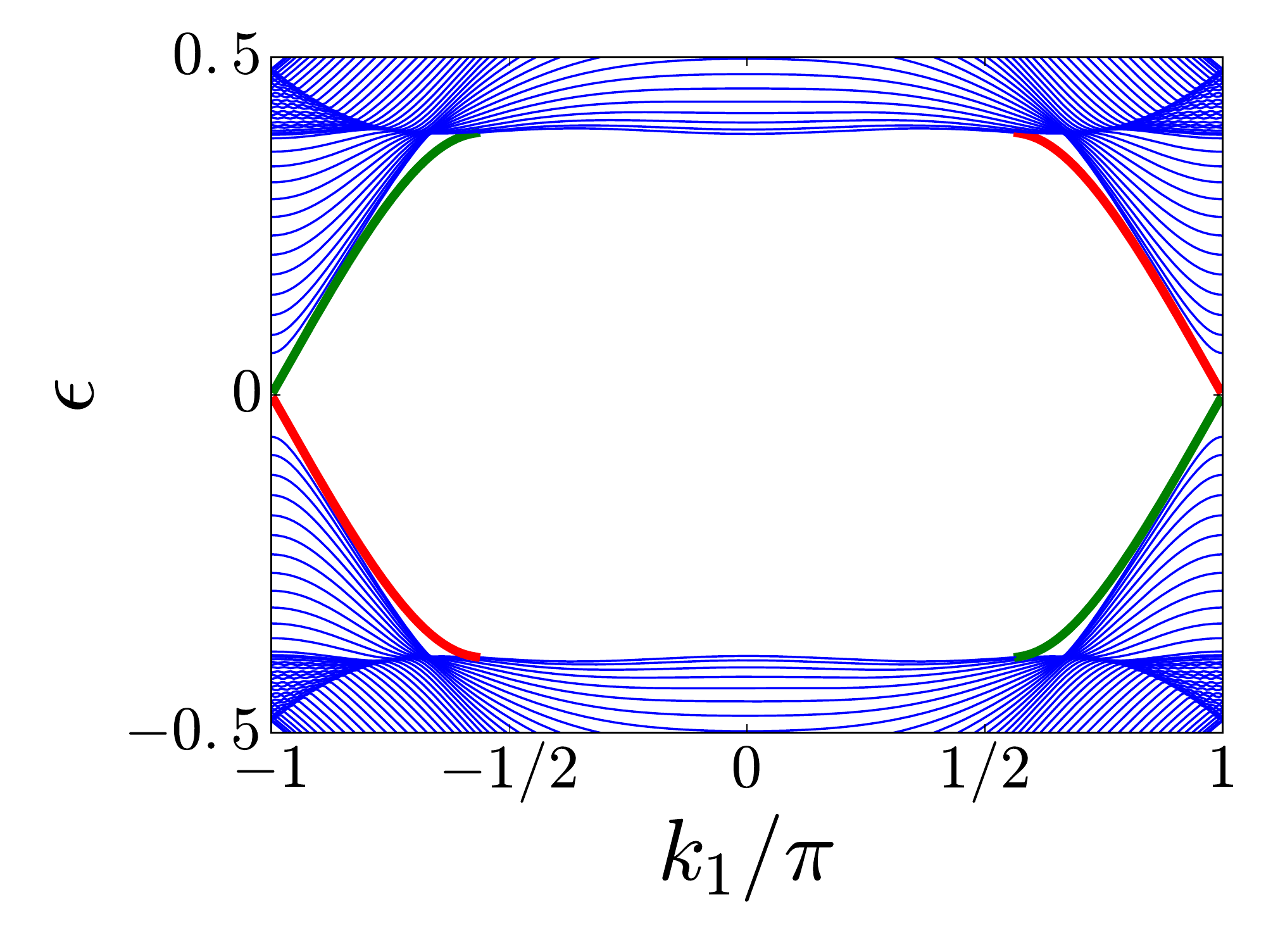}
}
\\
\centering
\subfloat[]{
\includegraphics[width=.4\textwidth]{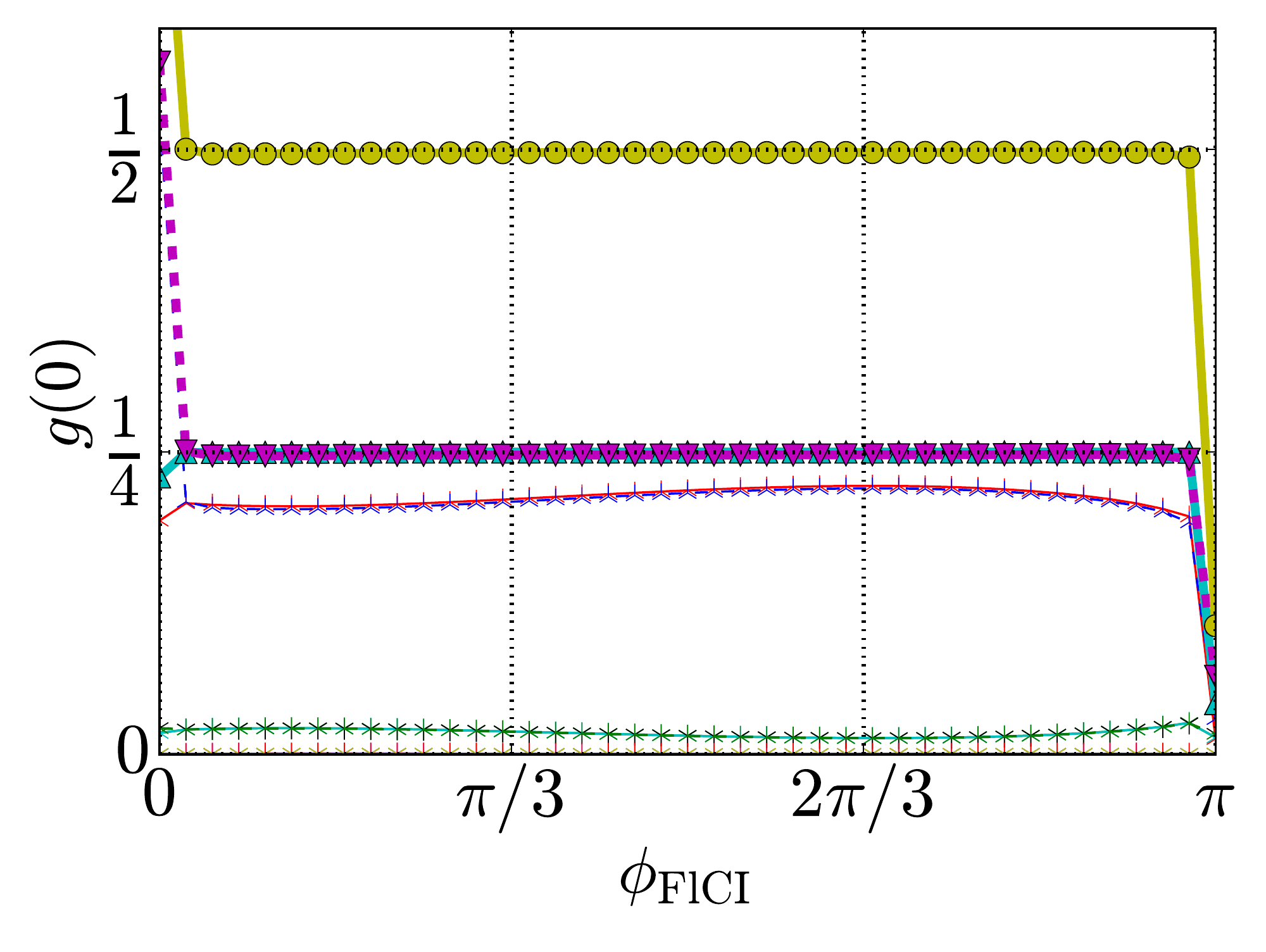}}
\subfloat[]{
\includegraphics[width=.4\textwidth]{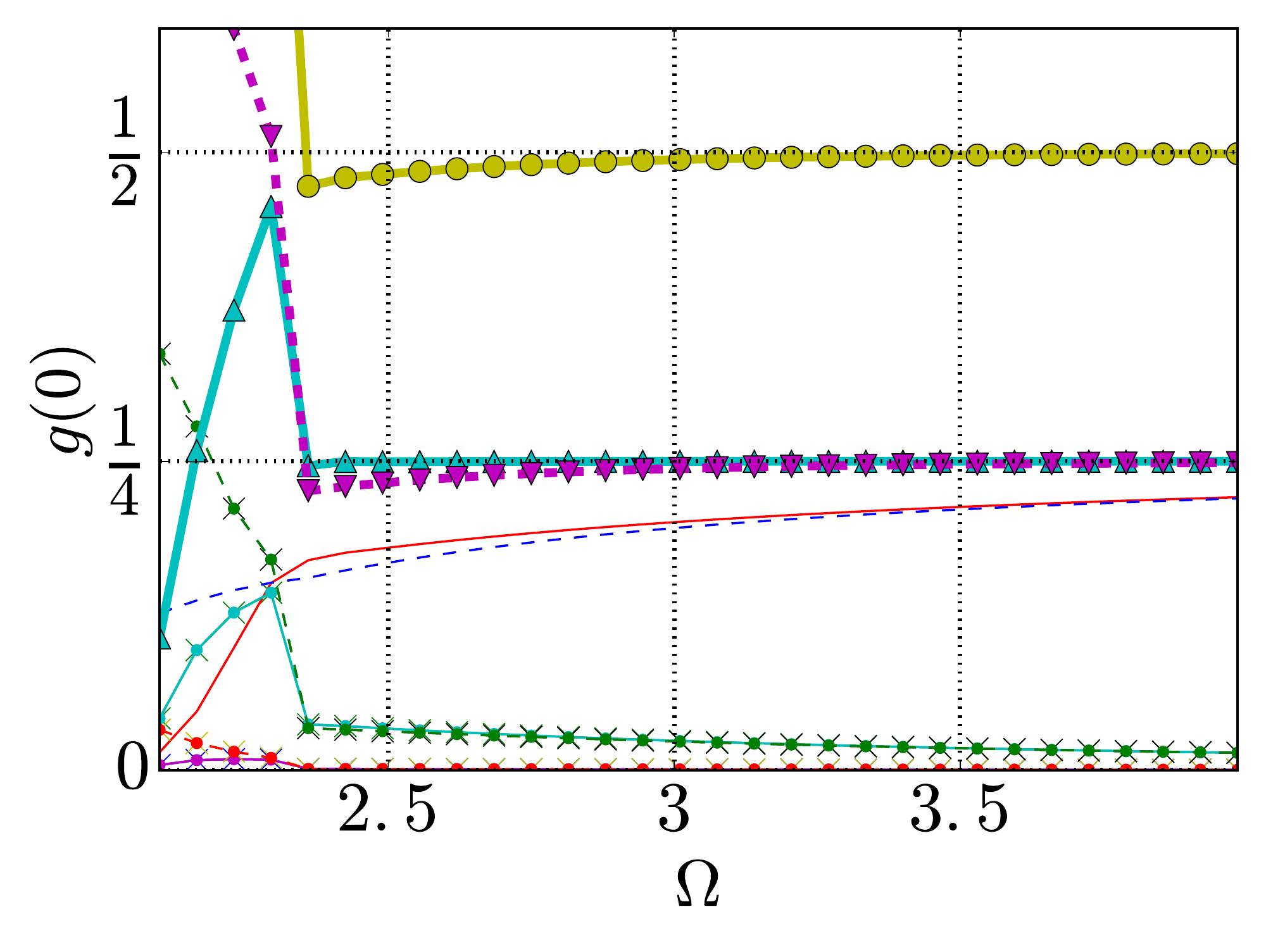}}
\\
\centering
\subfloat[]{
\includegraphics[width=.7\textwidth]{quenchedQXL_thickLegend.pdf}
}
\caption{Top panel: Spectra of (a): FlCI, (b): FTSC. Middle panel: DC conductance at $E_\mathrm{F}=0$, as functions of the (c) polarization angle at FlCIs, $\phi_\mathrm{FlCI}$ and the (d): drive frequency $\Omega$. Bottom panel (e): Legends of plots (c) and (d). Parameters: (a): $t_1=1,A=1.25,\Omega=3.4,\phi=0.9\pi$, (b): $t_1=1,\Delta=0.5,A=1.9,\Omega=3.4,\phi=\pi/2$. } 
\label{fig:graphene}
\end{figure*}

Thus we take $t_2=0$, so that Eq.~\eqref{eq:graphene} simply describes a graphene with nearest-neighbor hopping. The static device now consists of a graphene ribbon, where the center is proximitized by s-wave superconductor. Once periodic driving fields with appropriate parameters are added, we observe a $\mathcal{C}=1$ FlCI [Fig.\ref{fig:graphene}(a)] and $\mathcal{N}=1$ FTSC [Fig.\ref{fig:graphene}(b)].

Knowing that our Floquet device may yield a single FlCI fermion and a single CMEM {  at one} {edge} { of the system}, we proceed to probe the DC transport.

In Fig.\ref{fig:graphene}(c), the DC conductance is calculated as a function of the polarization angle at the FlCIs. We observe that within $[0,\pi]$, almost all angles, except those at the boundary values $0,\pi$, yield half-quantized DC conductance. This is because for $\phi_\mathrm{FlCI}=0,\pi$, the light is linearly polarized and does not break time-reversal symmetry\cite{Kitagawa}. Next, we vary frequency and find that beyond a critical value [near $\Omega=3$ in Fig.\ref{fig:graphene}(d)], one always obtains $1/4$ normal transmission and local Andreev reflection. In fact, extrapolating to the high-frequency regime and performing a Magnus expansion\cite{Kitagawa}, it is expected that one can always gap the normal-superconducting graphene to generate a chiral regular fermion and a CMEM, except that the $\pm1$ Floquet sideband contributions tend to zero asymptotically as $\Omega$ increases. 
\subsection{Higher half-quantized plateaus from a driven Haldane model} \label{sec:HaldaneContinuous}
As alluded to, Floquet systems boast a wide range of phases with large topological invariants\cite{JBGong1,Derek2,LWZhou2,TSXiong,Martin,OkaWeyl}. This feature is all the more appealing for the transport signature of CMEMs: While alternative mechanisms may explain half-quantized plateaus\cite{XGWen,JaySau}, it is unlikely that these mechanisms, of static origin, are able to induce \emph{higher} half-quantized plateaus from models with minimal number of bands. In addition, for the purpose of TQC, a larger number of MZMs allows more qubits to be encoded.

Therefore, we investigate the transport signature of devices harboring more than one CMEMs. To look for such phases, we choose the Haldane model with s-wave pairing potential, described by Eqs.~\eqref{eq:Hamiltonian} and \eqref{eq:graphene}. Experimentally, by means of ultracold fermionic gases, the former was realized using $^{40}$K atoms modulated in time\cite{HaldaneExperiment}, the latter via $^6$Li hyperfine state pairs close to Feshbach resonance\cite{swave}. The left and right parts are given by the Haldane model without s-wave pairing, and the entire device is irradiated by elliptically polarized light.
\begin{figure*}
\subfloat[]{
\includegraphics[width=.4\textwidth]{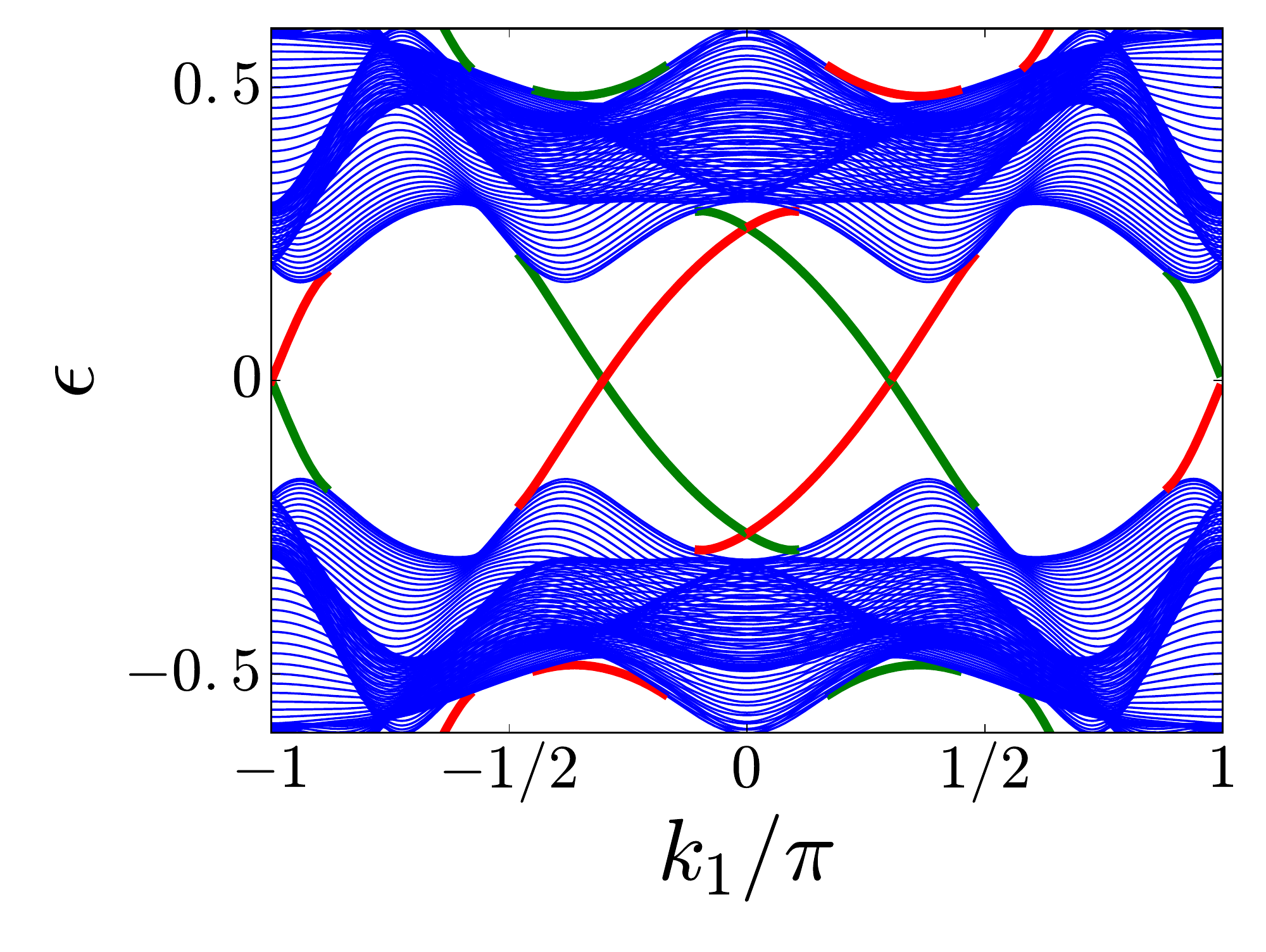}
}
\subfloat[]{
\includegraphics[width=.4\textwidth]{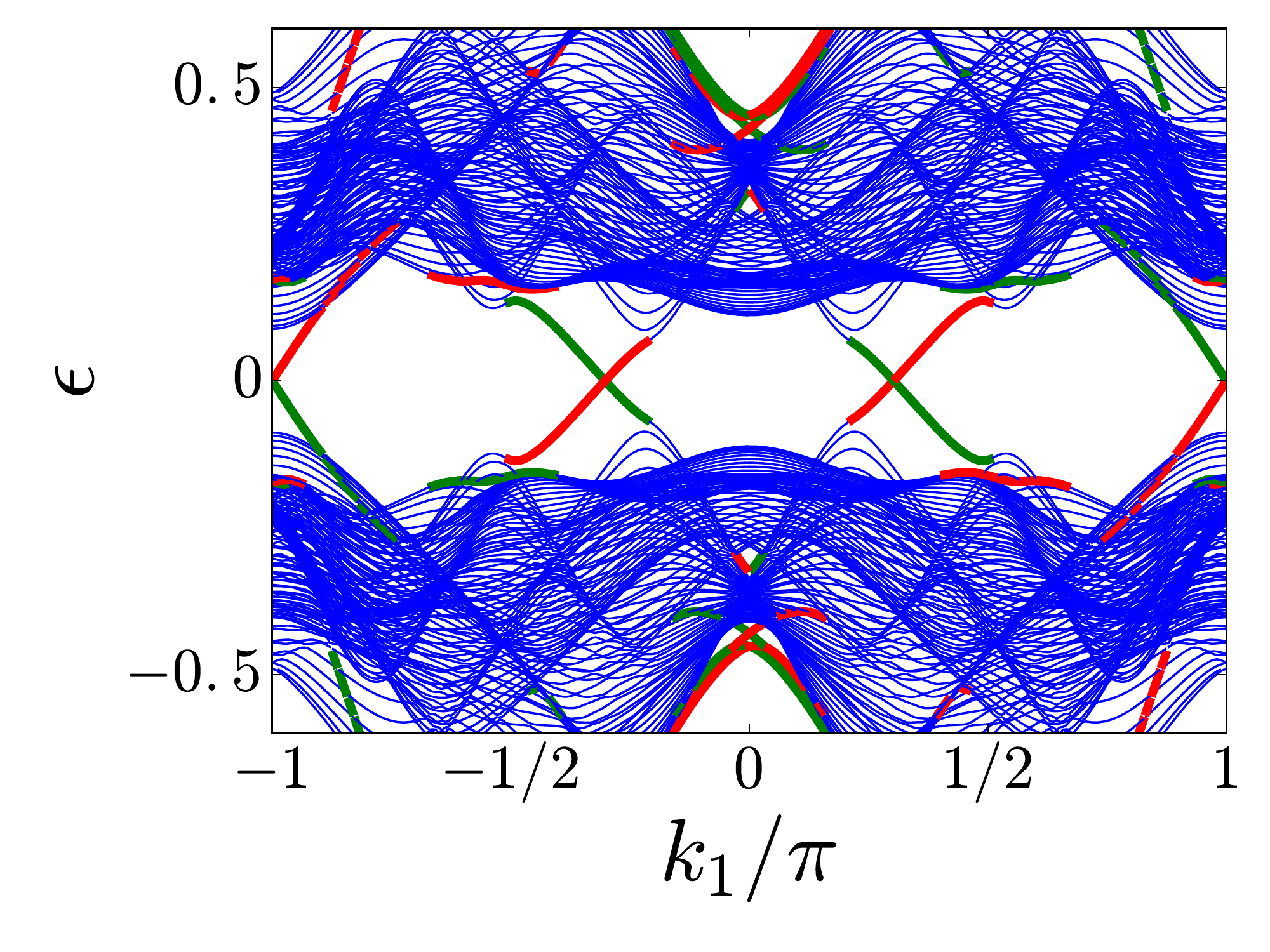}
}
\\
\subfloat[]{
\includegraphics[width=.4\textwidth]{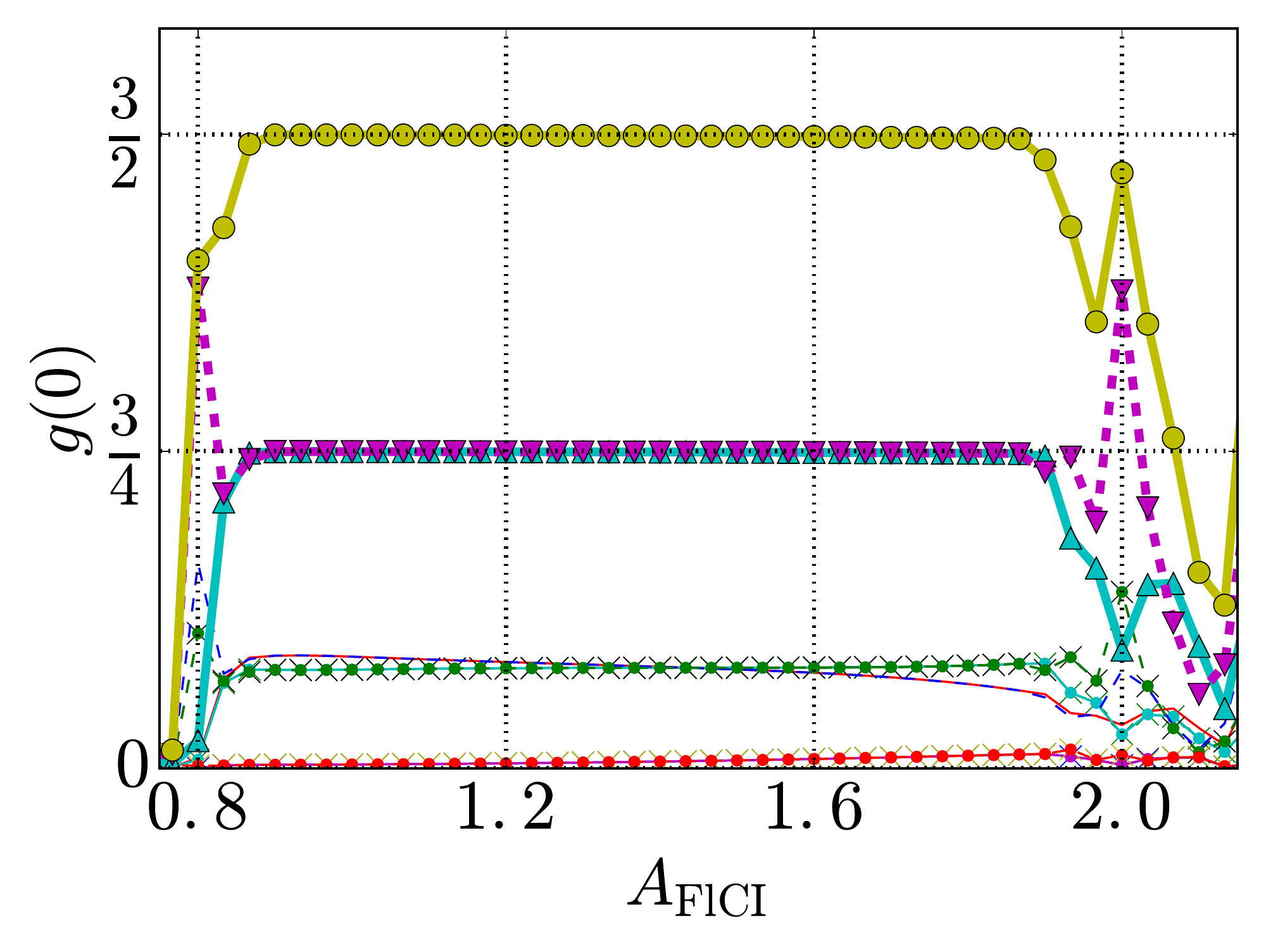}
}
\subfloat[]{
\includegraphics[width=.4\textwidth]{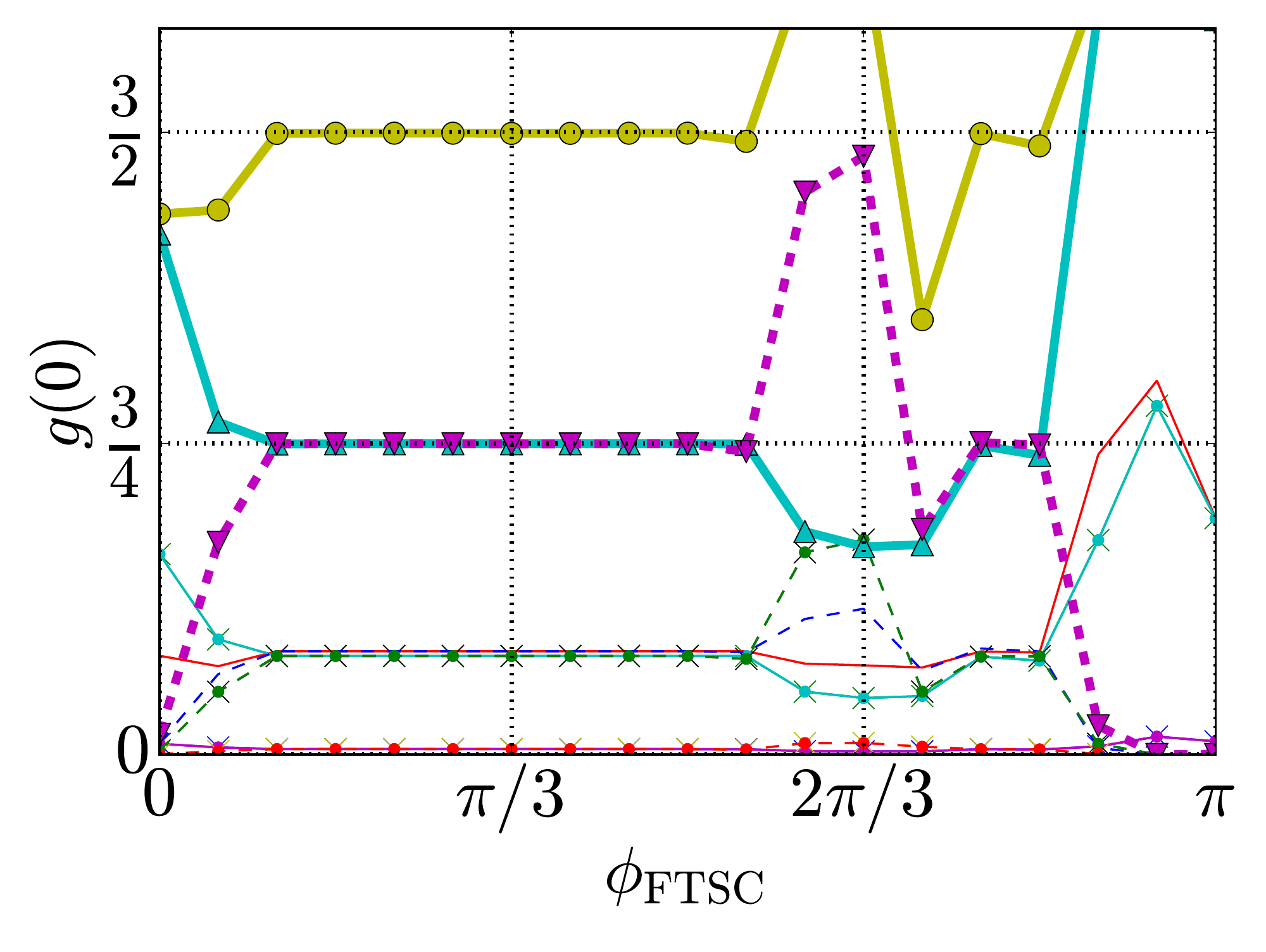}
}
\\
\centering
\subfloat[]{
\includegraphics[width=.7\textwidth]{quenchedQXL_thickLegend.pdf}
}
\caption{Top panel: Spectra of (a): FlCI, (b): FTSC. Middle panel: DC conductance at $E_\mathrm{F}=0$ as functions of the (c): light intensity at the FlCIs and the (d): polarization angle at the FTSC, $\phi_\mathrm{TSC}$. Bottom panel (e): legends of plots for (c) and (d). Parameters: (a): $t_1=0.8,t_2=0.3,\Phi=\pi/2,A=1.25,\Omega=1.5,\phi=\pi/2$, (b): $t_1=1,t_2=0.45,\Phi=\pi/2,\Delta=0.6,A=1.75,\Omega=1.5,\phi=\pi/3$.}
\label{fig:CPLHaldane}
\end{figure*}

We now discuss the anticipated transport behavior when several CMEMs are present. For the case of a single CMEM, the splitting of a chiral ordinary fermion into two CMEMs (see Sec.~\ref{sec:HalfQuantized} or Refs.\cite{SCZhang3,SCZhang4}) is a major reason behind its half-quantized conductance. Now suppose the device hosts $2n+1$ chiral ordinary edge modes and $2n+1$ CMEMs, with $n>0$. Guided by Occam's razor, an immediate intuition is that each of the $2n+1$ incoming chiral modes will undergo its own splitting, giving rise to $1/4$ local Andreev reflection and normal transmission each, so that the conductance will be given by $(2n+1)/2$.

To verify our insight above, considering that the next entry in the hierarchy of half-quantized conductance is $3/2$, we seek phases with three edge modes in both the FlCIs and the FTSC. An example is given in Figs.~\ref{fig:CPLHaldane}(a-b), where the three chiral edge modes of interest occur at the central Floquet gaps. Then, we compute the DC conductance at $E_\mathrm{F}=0$ by varying the light intensity at the FlCIs [Fig.~\ref{fig:CPLHaldane}(c)] and the polarization angle at the FTSC [Fig.~\ref{fig:CPLHaldane}(d)]. {In Fig.~\ref{fig:CPLHaldane}(c)}, the $n=\pm1$ sidebands already constitute plateaus, with the $n=0,\pm2$ sidebands drifting not too far away from their respective values.   {In Fig.~\ref{fig:CPLHaldane}(d)}, all sidebands form non-quantized plateaus themselves. While the exact conductance values may not bear any discernible feature, the sidebands again always have equal normal transmission and local Andreev reflection. Furthermore, when they are summed over, we obtain $3/4$ plateaus $(\bar{g},\bar{g}^\mathrm{LAR})$, which in turns yield $3/2$ for the total longitudinal DC conductance. This result strongly supports our earlier intuition that for multiple incoming chiral fermions, if the FTSC permits formation of CMEMs with the same number and chiralities, then equal scattering processes for each incoming mode is favored, and higher half-quantized plateaus ensue.


\subsection{$\pi$ Floquet CMEMs and their transport}\label{sec:threehalfQuenched}
\begin{figure*}
\subfloat[]{
\includegraphics[width=.4\textwidth]{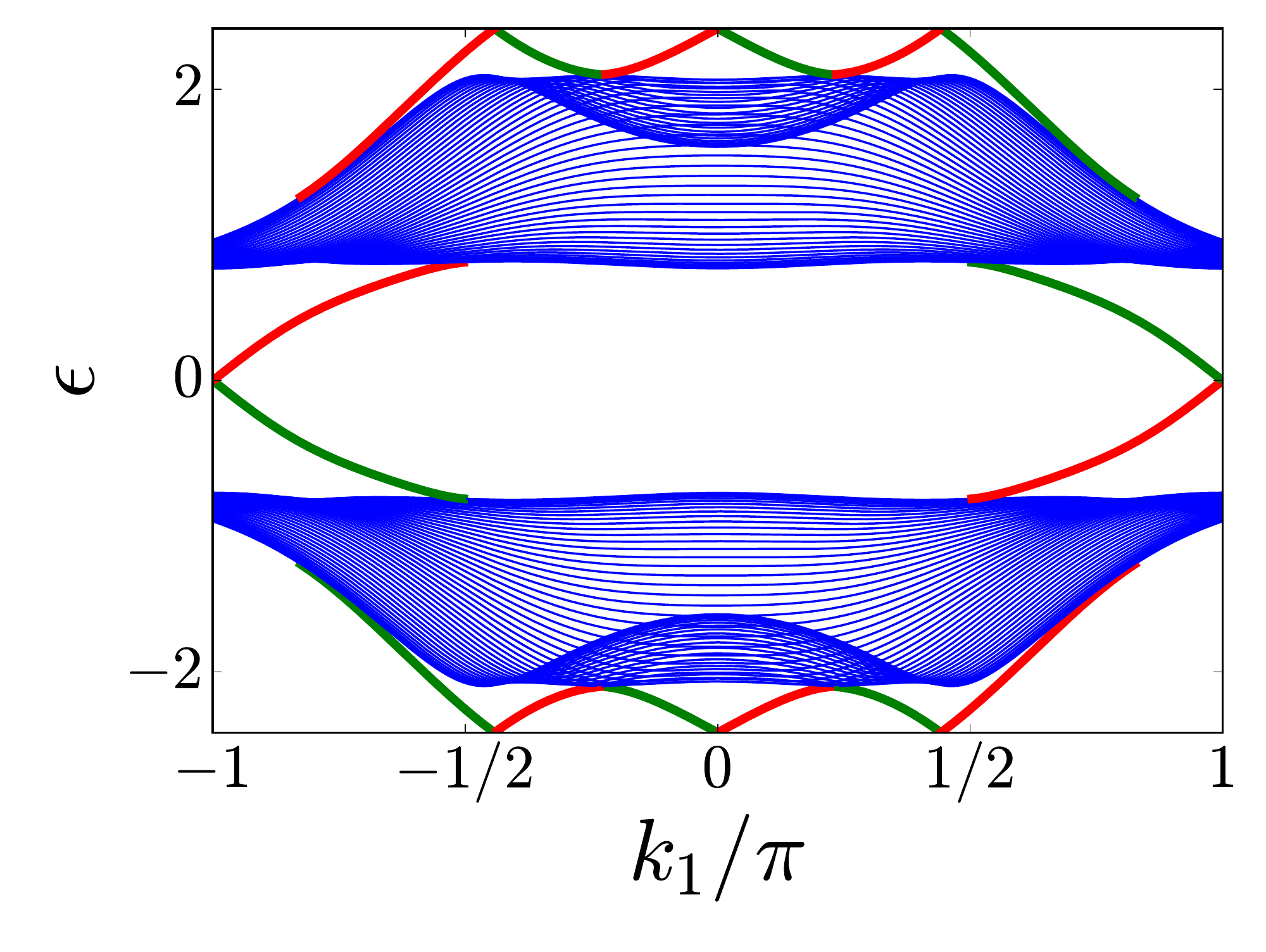}
}
\subfloat[]{
\includegraphics[width=.4\textwidth]{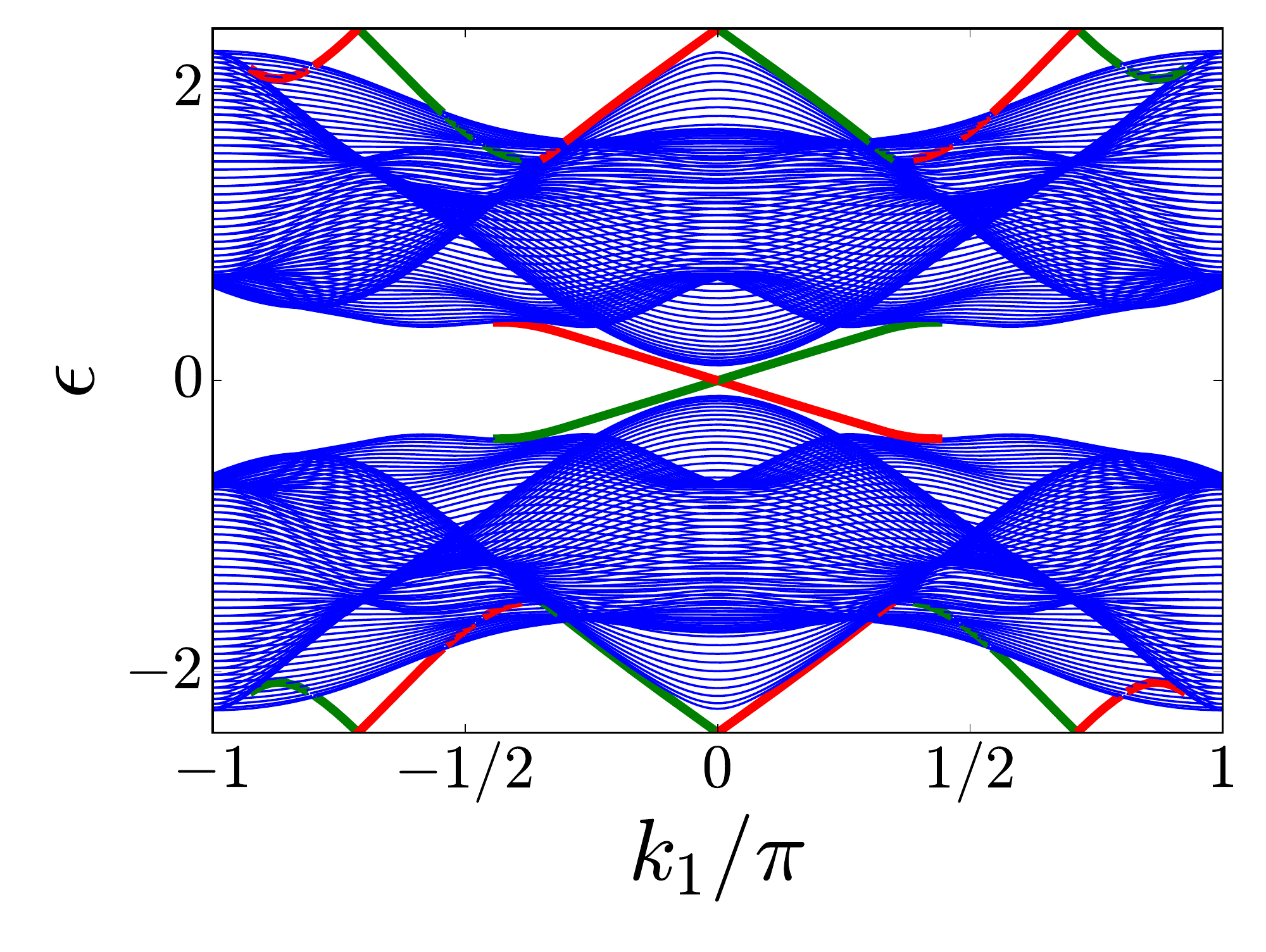}
}
\\
\subfloat[]{
\includegraphics[width=.4\textwidth]{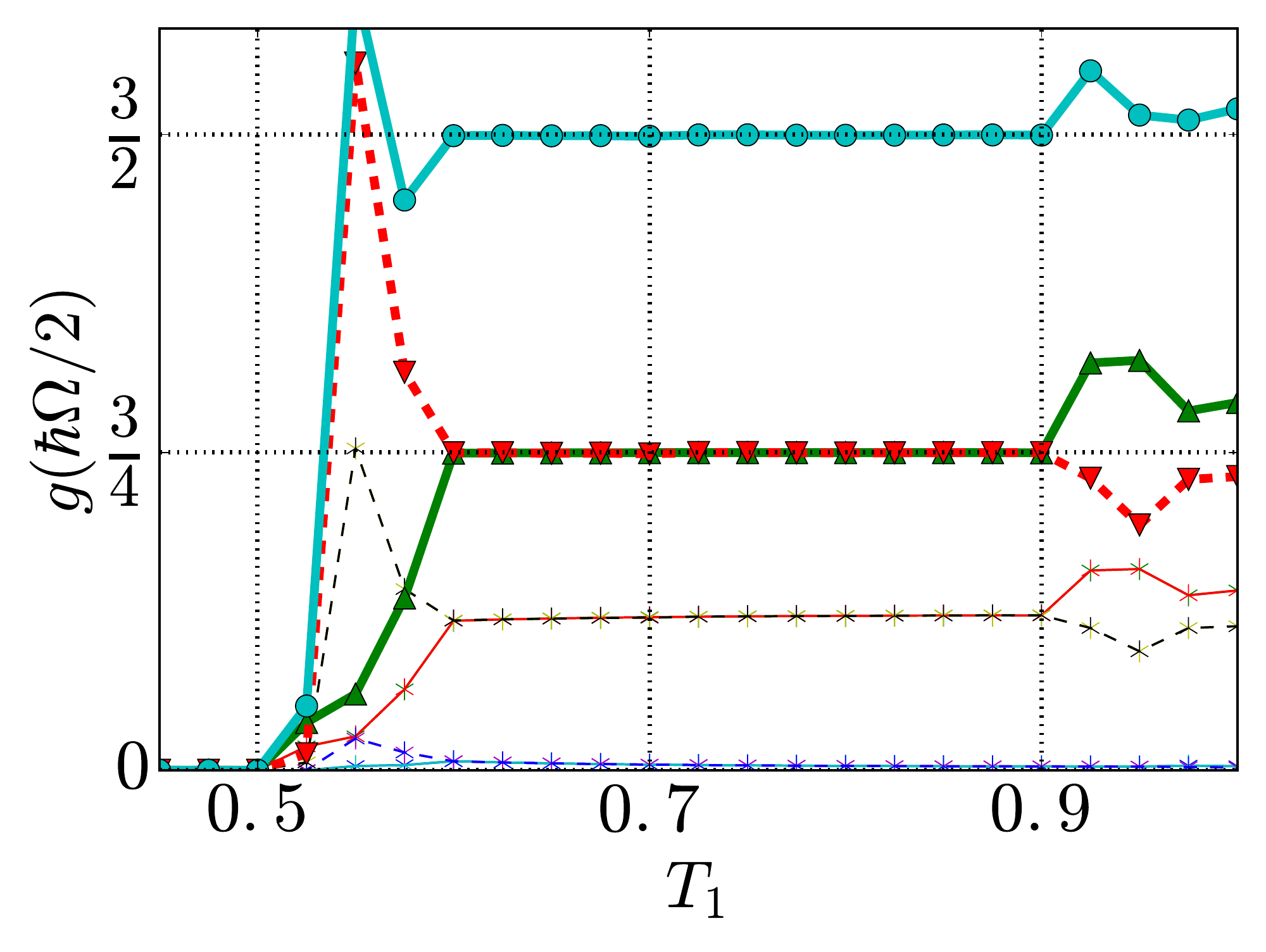}
}
\subfloat[]{
\includegraphics[width=.4\textwidth]{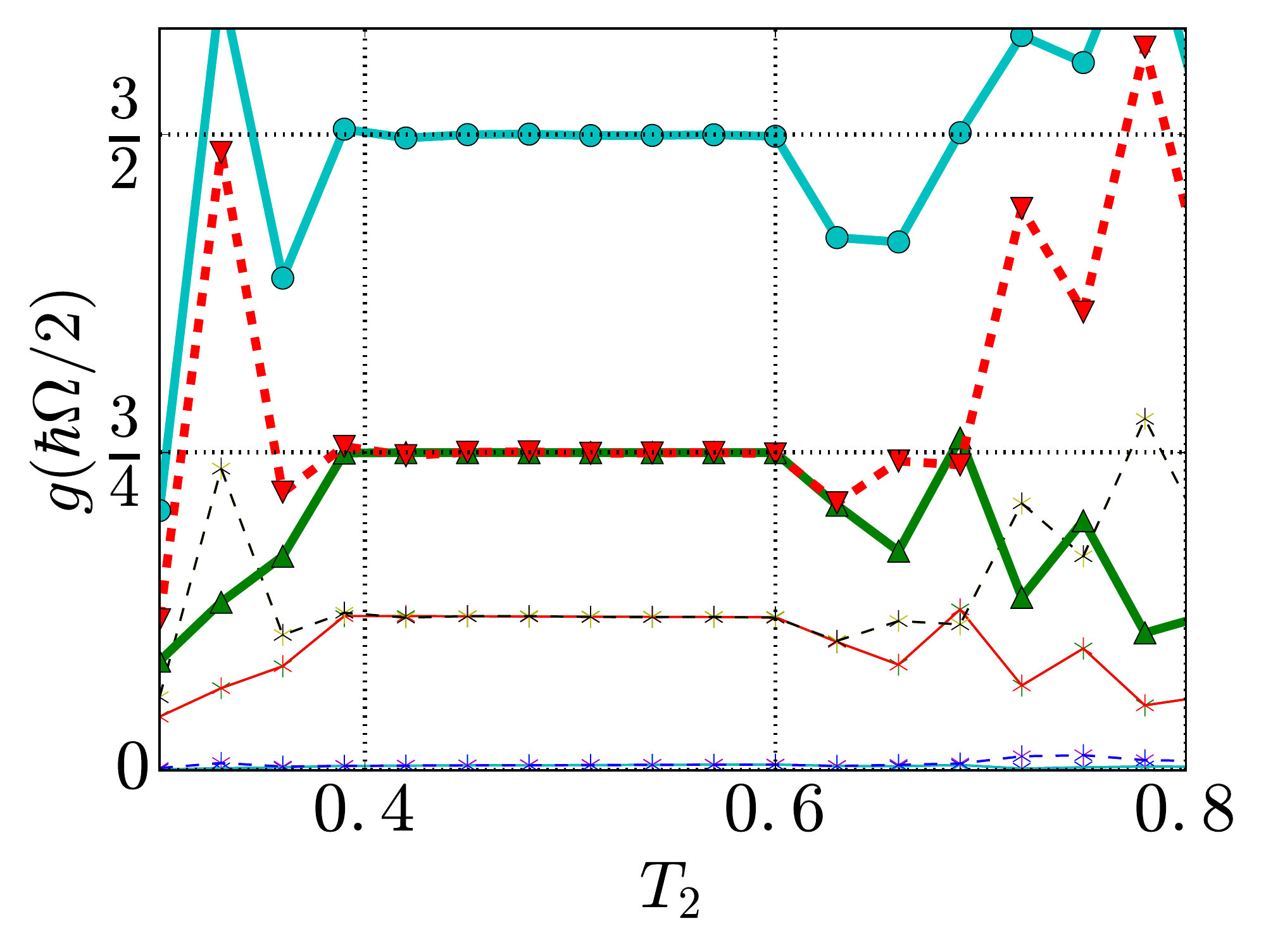}
}
\\
\centering
\subfloat[]{
\includegraphics[width=.7\textwidth]{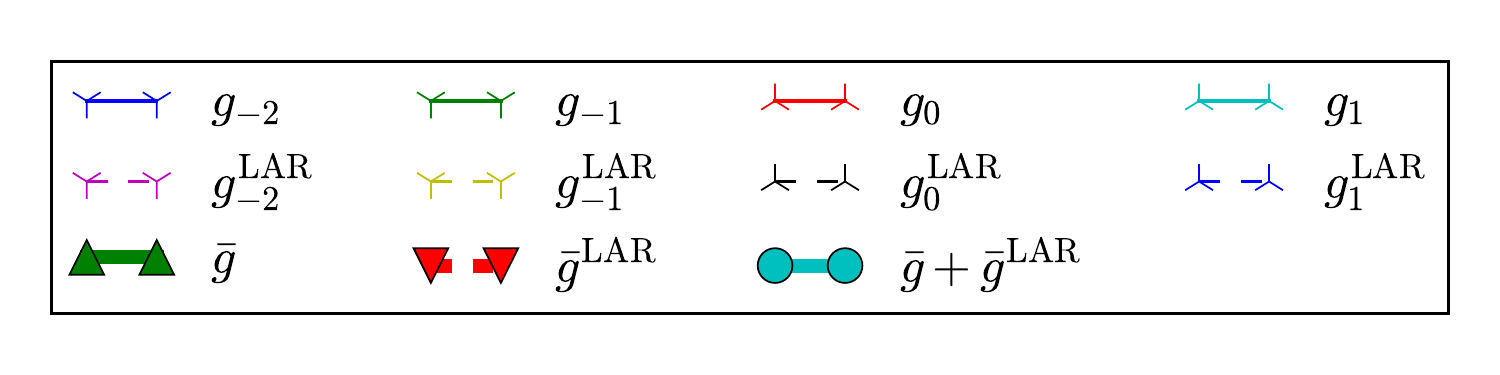}
}
\caption{Top panel: Spectra of (a): FlCI, (b): FTSC. Middle panel: DC conductance at $E_\mathrm{F}=\hbar\Omega/2$ as the quench durations (c) $T_1$ and (d) $T_2$ vary. Bottom panel (e): Legends of plots (c) and (d). Parameters: $t_1=1,t_2=0.8,t_3^{(1)}=1,t_3^{(2)}=-1,\Delta=1.5,\Phi^{(1)}=\pi/6,\Phi^{(2)}=-\pi/2,T_1=0.7,T_2=0.6$, applicable to both (a) and (b) except the pairing $\Delta$ which is present only at the FTSC (b).}
\label{fig:Pi}
\end{figure*}
With a glimpse of the transport properties of multiple CMEMs, we are now ready to address the case of CMEMs that can only exist in Floquet topological phases, namely, those crossing $\epsilon=\hbar\Omega/2$ in Floquet spectra.   {Because quasienergies $\pm\hbar\Omega/2$ correspond to eigenphases $\pm\pi$, they are also known as Floquet $\pi$ modes\cite{Raditya}. Many peculiar features of Floquet systems owe their existences to these $\pi$ modes}, such as non-adiabatic quantized pump\cite{Titum}, novel bulk-edge correspondence\cite{Kitagawa2}, band holonomy\cite{LWZhou1}, etc. In the quest for Majorana fermions, the fact that quasienergies $\pm\hbar\Omega/2$ are identical implies a larger search space for their realizations. While this is generally recognized\cite{JBGong1,FloquetPi,FloquetMajorana1}, studies on their transport signature are scarce\cite{Kundu}.

  {
To look for Floquet $\pi$ modes, we consider a quenched Haldane model with third-neighbor hoppings by adding an extra term to} Eq.~\eqref{eq:graphene}: $h(\bm{k})\mapsto h(\bm{k})+V(\bm{k})$, where:
\begin{equation}
\begin{split}
V(\bm{k})&=t_3\{2\cos[\bm{k}\cdot(\bm{a}_1-\bm{a}_2)]+\cos[\bm{k}\cdot(\bm{a}_1+\bm{a}_2)]\}\sigma_x \\
&+t_3 \sin[\bm{k}\cdot(\bm{a}_1+\bm{a}_2)]\sigma_y.
\end{split}
\end{equation}
Then, we periodically quench the parameters $\Phi,t_3$ to obtain Floquet spectra as illustrated in Figs.~\ref{fig:Pi}(a-b), where the gaps at $\epsilon=\pm\hbar\Omega/2$ harbor three $\pi$ modes for both the FlCIs and the FTSC. 

Thus, we study the DC conductance at $E_\mathrm{F}=\hbar\Omega/2$, as functions of quench durations $T_{1,2}$. The explanation for $\frac{3}{2}\frac{e^2}{h}$ DC conductance in Sec.\ref{sec:HaldaneContinuous} carries over, because both the number and chirality of the incoming FlCI chiral fermions and CMEMs are the same, and $\epsilon=\hbar\Omega/2$ is degenerate with $\epsilon=-\hbar\Omega/2$.

Notice however that for Floquet CMEMs at $\epsilon=0$, the sideband contributions are symmetrically distributed around a single $n=0$ sideband, so that there is an odd number of dominant sidebands [$n=\pm1$ around $n=0$ in Figs.\ref{fig:graphene}-\ref{fig:CPLHaldane}(c-d)]. On the contrary, for the $\epsilon=\hbar\Omega/2$ CMEM, the non-negligible sidebands are distributed around a \emph{twin} central sidebands [$n=-1,0$ in Fig.\ref{fig:Pi}(c-d)], i.e. there is an even number of dominant sidebands. This observation is consistent with previously-reported results in Floquet quantum Hall insulators\cite{HHYap}, and may serve as a telltale sign of distinguishing $\pi$ and zero CMEMs in Floquet systems.

Our example for Floquet $\pi$ CMEMs also illustrates another distinctive feature of Floquet topological phases. For static junctions, by the bulk-edge correspondence, it may be possible\footnote{Two recent works\cite{Bena,MajoranaBulkEdge} gave examples where the Chern numbers no longer correspond to the number of MBS.} to deduce the transport signature from the band Chern numbers of the CIs and TSC\cite{SCZhang4}. Consider now the parameters for the Floquet device as described in the caption of Fig.~\ref{fig:Pi}, which yields the following Floquet-Chern numbers: $\mathcal{C}=2$ for the FlCIs, and $\mathcal{N}=4$ for the FTSC. If it were a static system, the longitudinal conductance could be $\frac{2e^2}{h}$, due to transmission alone, without normal nor crossed Andreev reflection. As discussed, there actually are three Floquet chiral ordinary fermions and CMEMs, which yield higher half-integer DC conductance of $\frac{3}{2}\frac{e^2}{h}$. This is because the quasienergy is defined modulo drive frequency: $\epsilon=\epsilon \mod{\hbar\Omega}$, so that it is possible for edge states to wind across the Floquet-Brillouin zone\cite{Lindner2}, implying that the Floquet-Chern number is no longer in direct correspondence with the number of chiral edge modes. Hence, in the study of Floquet junctions, one should use topological invariants based on Floquet gaps\cite{Lindner2} which reflect more closely the number of edge modes.

\subsection{Disorder}\label{sec:Disorder}
In the first TI-based experimental realization of CMEM\cite{SCZhangExperiment}, half-quantized plateaus were observed during sweeps of magnetic field. While this induces topological phase transitions for a single CMEM\cite{SCZhang3} to emerge, it also generates many magnetic domains\cite{SCZhangReply,JaySau,XGWen,KTLaw}, resulting in a highly-disordered sample. Hence, based on the theory of percolation transition, several authors have proposed alternative mechanisms\cite{XGWen,JaySau} that can give rise to half-quantized conductance in QAHI-TSC-QAHI heterostructures without CMEMs.
\begin{figure*}
\subfloat[]{
\includegraphics[width=.4\textwidth]{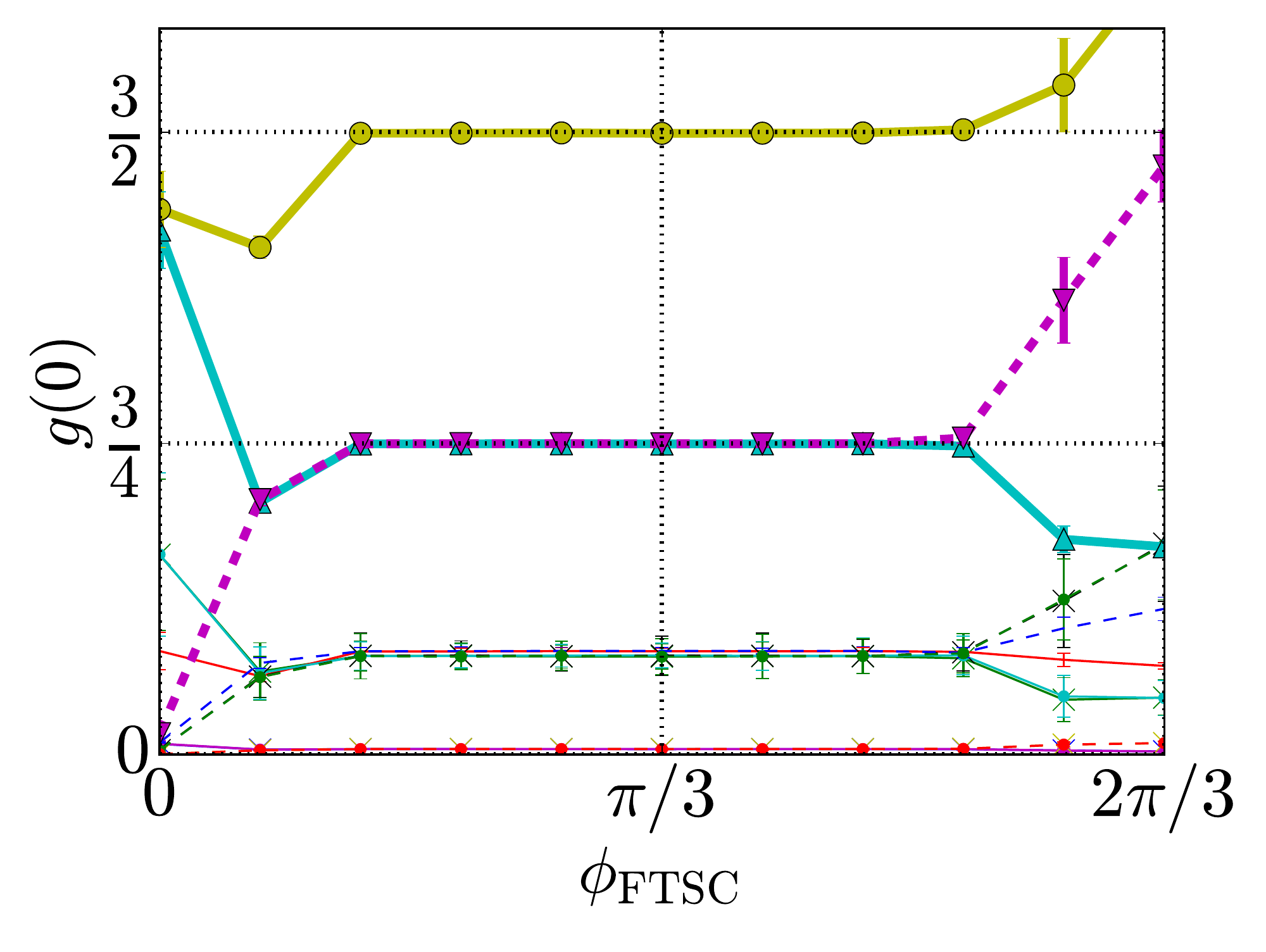}
}
\subfloat[]{
\includegraphics[width=.4\textwidth]{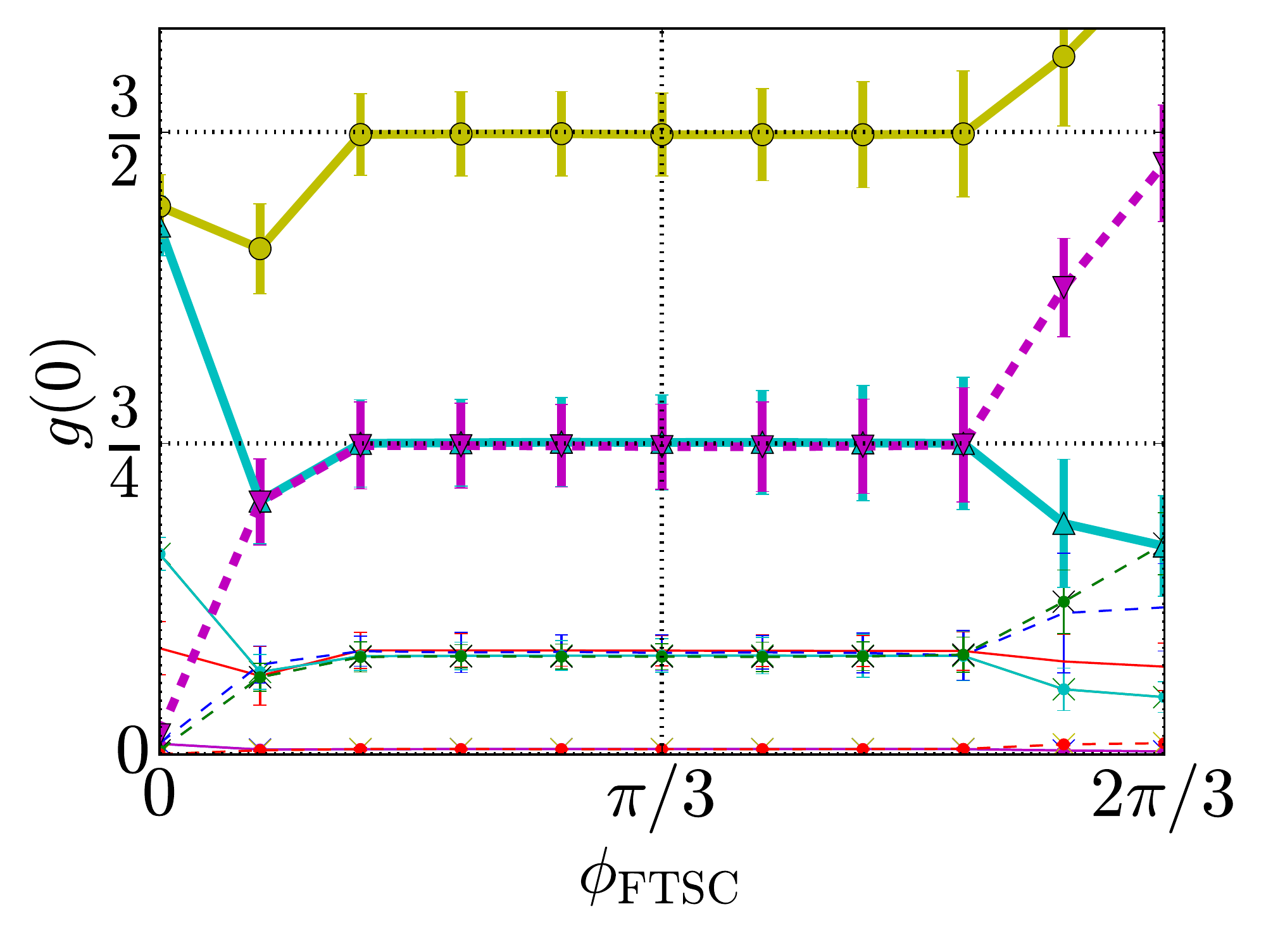}
}
\\
\subfloat[]{
\centering
\includegraphics[width=.7\textwidth]{quenchedQXL_thickLegend.pdf}
}
\caption{DC conductance of the model in Sec.~\ref{sec:HaldaneContinuous}, evaluated at $\epsilon_\mathrm{F}=0$ with random onsite potentials. (a): Static disorder. (b): Time-periodic disorder at frequency $\Omega$. The disorder is uniform in sublattice space and has a box-like distribution in $[-W,W]$, where $W\approx 0.4\delta$, with $\delta$ the size of the central Floquet gap in Fig.~\ref{fig:CPLHaldane}(b). The error bars are magnified by three times, and represent the standard deviation of 45 disorder realizations.} 
\label{fig:disorderZero}
\end{figure*}
\begin{figure*}
\subfloat[]{
\includegraphics[width=.4\textwidth]{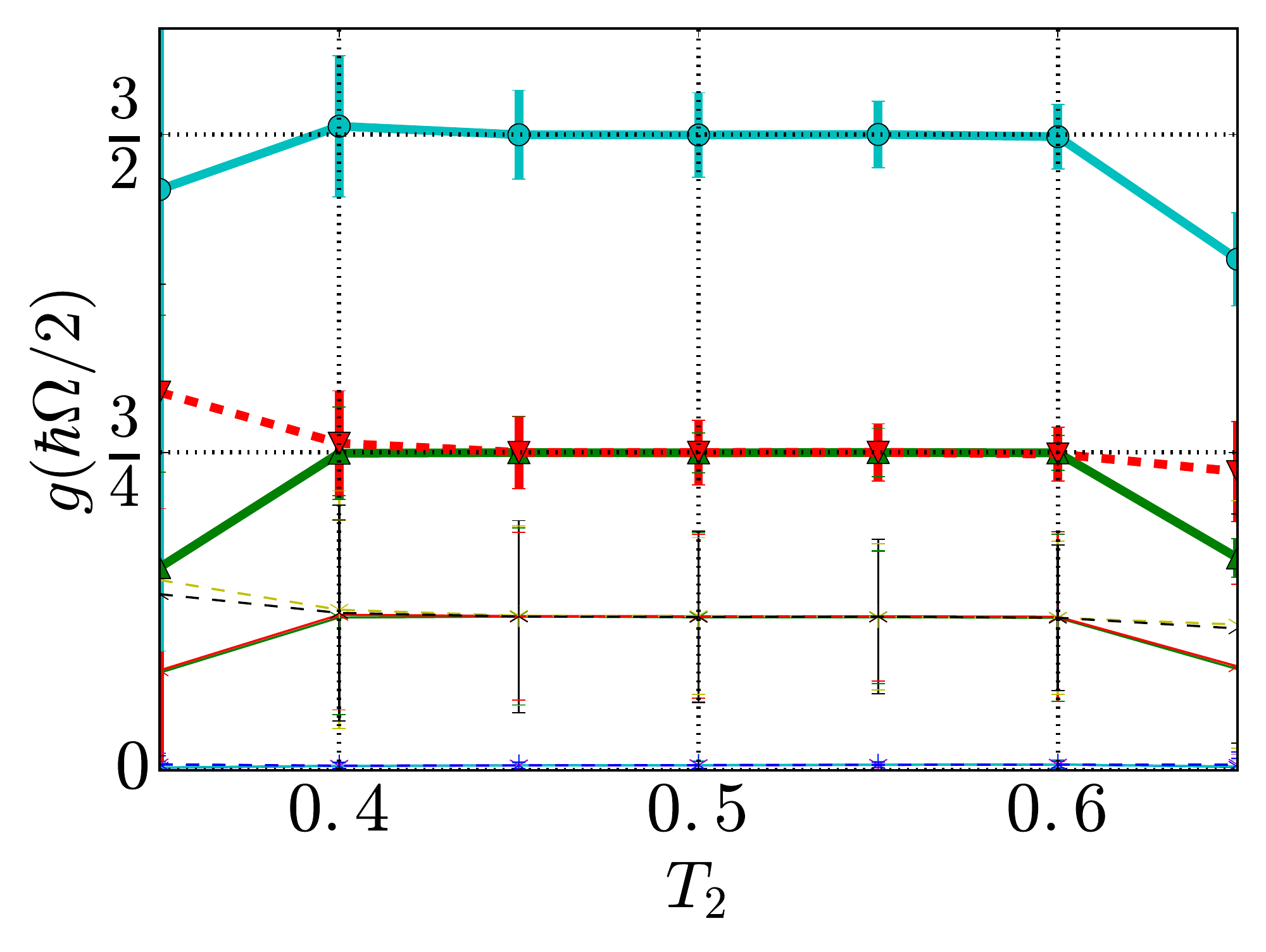}
}
\subfloat[]{
\includegraphics[width=.4\textwidth]{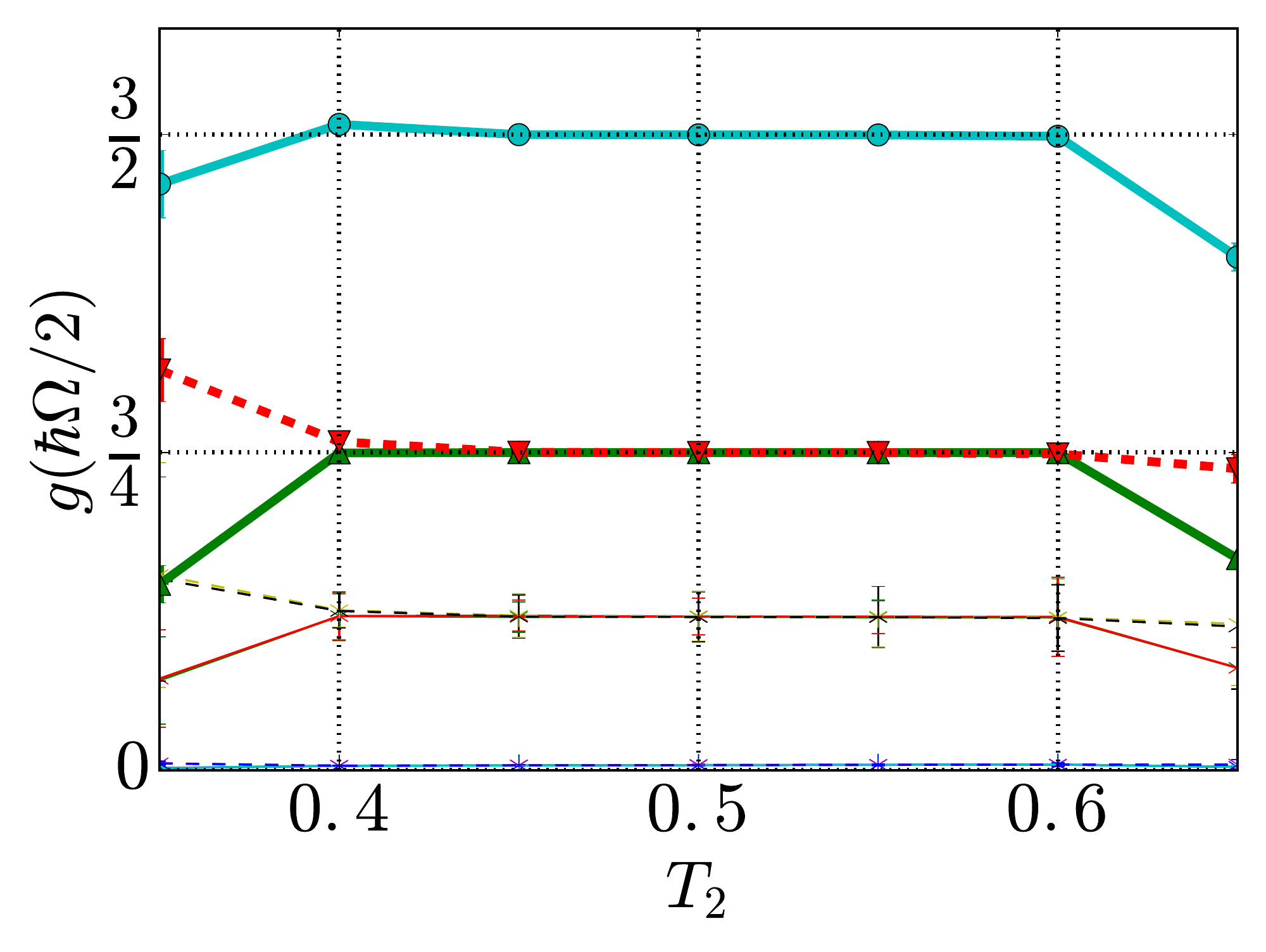}
}
\\
\centering
\subfloat[]{
\includegraphics[width=.7\textwidth]{Pi_thickLegend.pdf}
}
\caption{DC conductance of the model in Sec.~\ref{sec:threehalfQuenched}, evaluated at $\epsilon_\mathrm{F}=\hbar\Omega/2$ with random onsite potentials. (a): Static disorder. (b): Time-periodic disorder at frequency $\Omega$. The disorder is uniform in sublattice space and has a box-like distribution in $[-W,W]$, where $W\approx 0.4\delta$, with $\delta$ the size of the $\pm1$ Floquet gap in Fig.\ref{fig:Pi}(b). The error bars are magnified by 10 times, and represent the standard deviation of 45 disorder realizations.} 
\label{fig:disorderPi}
\end{figure*}

Considering the prominent roles that disorder may play, here we ask whether the higher half-quantized plateaus due to multiple Floquet CMEMs can survive disorder. To that end, we model disorders as random onsite potentials, uniform in spin or sublattice space, and distributed uniformly in $[-W,W]$. We consider disordered samples from Secs.\ref{sec:HaldaneContinuous},\ref{sec:threehalfQuenched}, and compute the DC conductance for different Floquet sidebands at $\epsilon_\mathrm{F}=0$. In Floquet systems, one may incorporate the effect of time-dependent disorder, provided the time dependence is periodic at the same fundamental frequency of the driving field.

We observe the topological protection of the $1/4$ splitting of the three Floquet CMEMs against disorder of size $\approx 0.4$ the superconducting gap in Fig.~\ref{fig:CPLHaldane}(b). Interestingly, for static disorder, the $3/4$ plateaus for both normal transmission and local Andreev reflection are robust at each realization. On the other hand, the system appears to be less resistant against time-periodic disorder, since it is only after averaging that the $3/4$ plateaus remain stable.

Next, we probe the robustness of the $\pi$ Floquet CMEMs Sec.\ref{sec:threehalfQuenched}. For the sake of comparison, we similarly choose a disorder strength $\approx 0.4$ the size of the $\pm\pi$ Floquet gap in spectrum \ref{fig:Pi}(b). Remarkably, the robustness against disorder is swapped as compared with the zero modes considered previously. More precisely, the $\pi$ Floquet modes are more robust to time-periodic disorder, less so to static disorder, whereas for the zero CMEMs it is the reverse. In total, the $\pi$ CMEMs from Sec.\ref{sec:threehalfQuenched} are more resistant against disorder compared with their zero analogs from Sec.\ref{sec:HaldaneContinuous}, since the errorbars are magnified by 10 and 3 times respectively [Figs.\ref{fig:disorderPi},\ref{fig:disorderZero}].

\section{Discussions}\label{sec:Discussions}
\subsection{On chiral Majorana edge modes}
Several authors\cite{SarmaQuantumComputation,XGWen,StanescuBook} have cautioned against the misuse of terminologies such as ``Majorana zero modes'' and ``Majorana fermions''. Thus we deem it necessary to explain in what context we are referring to the edge states in our systems as CMEMs.

By definition, superconductors always preserve the particle-hole symmetry, so that in two dimension with mixed boundary condition, eigenstates $\Psi_k$ at energy $E_k$ always have charge-conjugation partners $C\Psi^*_{-k}$ at energy $E_{-k}$\cite{Sato,Elliott,Chamon}, where $C=C^*=C^{-1}$ is the unitary part of the particle-hole symmetry. Thus, if the superconductor is gapped and yields zero or $\pi$ quasienergy solutions $\Psi_k$, one may construct an eigenstate $\tilde{\Psi}_k=\Psi_k+C\Psi^*_{-k}$ which obeys the Majorana condition $C\tilde{\Psi}_{-k}=\tilde{\Psi}_k$. It is in this regard that we refer to the gapless modes in the FTSC spectra [Figs.\ref{fig:graphene}-\ref{fig:Pi}(b)] as CMEMs.

%

\subsection{Origin of half-quantized conductance}\label{sec:HalfQuantized}
Quasiparticles with Majorana property are charge neutral\cite{Beenakker,Elliott}, and one may wonder how can any electrical conductance demonstrate the existence of such excitations. Indeed, the observation of half-quantized conductance in Ref.\cite{SCZhangExperiment} relies not on ``Majorana current'', but the following facts\cite{SCZhang3,SCZhang4}: (i) domain wall forms as CMEM between a $\mathcal{C}=1$ QAHI and a $\mathcal{N}=1$ TSC, and (ii) a CMEM is an equal-weight superposition of electron and hole. Therefore, an incoming chiral regular fermion from the QAHI can be scattered into two CMEMs: $\gamma_a$ on the interface between the QAHI and the TSC, or $\gamma_b$ on the interface between the TSC and vacuum. Now, $\gamma_a$ corresponds to normal reflection and local Andreev reflection, whereas $\gamma_b$ corresponds to normal transmission and crossed Andreev reflection. These processes must occur with equal probabilities $(=1/4)$, because $\gamma_{a,b}$ being Majorana fermions, are linear combinations of electron and hole with equal weights.

Next, working within the Anantram-Datta generalization\cite{DattaSuperconductor} of the Landauer-B\"{u}ttiker formalism\cite{Buttiker}, we model the sample as a three-terminal device, the three terminals being left, right, and superconductor, at voltages $V_\mathrm{L},V_\mathrm{R},V_\mathrm{sc}$ respectively. If the following conditions are met: (i) the superconductor is grounded, $V_\mathrm{sc}=0$, (ii) the left and right leads are biased symmetrically, $V_\mathrm{L}=-V_\mathrm{R}$, and (iii) all normal and Andreev processes have equal probabilities, then one can show that\cite{SCZhang4}, no current flows in or out the superconductor, and the only current flowing in the device is between the leads, with $I_\mathrm{L}=-I_\mathrm{R}=e^2V_\mathrm{L}/h$. In this case, the longitudinal conductance $g_{\mathrm{LR}}=I_\mathrm{L}/(V_\mathrm{L}-V_\mathrm{R})$ yields the half-integer $\frac{1}{2}\frac{e^2}{h}$.

This justifies using the two-lead setting as discussed in Sec.~\ref{sec:Method} to study the transport of such a device: Eq.~\eqref{eq:current} is left-right symmetric, namely, the current leaving the left lead equals the current entering the right lead. Thus even if there is superconductivity in the sample, it draws no current, so that we are exactly in the configuration as described in the previous paragraph, and the conductance in Eq.~\ref{eq:differentialConductance} is equivalent to the one used in Refs.~\cite{SCZhang3,SCZhang4}.

Importantly, Eq.~\eqref{eq:current} has the advantage that the normal transmission $g$ and local Andreev reflection $g^\mathrm{LAR}$ can be calculated separately. In fact, even the crossed Andreev reflection $g^\mathrm{CAR}$ can be computed. Hence, at least numerically, when conductance of $\frac{1}{2}\frac{e^2}{h}$ are found, one can inspect these scattering coefficients individually and substantiate the claim that CMEMs give rise to equal probability in all scattering processes, if all three of ${g}$, ${g}^\mathrm{LAR}$ and ${g}^\mathrm{CAR}$ are given by $1/4$. This is indeed what we find, but in order not to encumber the already saturated figures, we do not display any results for $g^\mathrm{CAR}$.

\begin{figure}
\includegraphics[width=.4\textwidth]{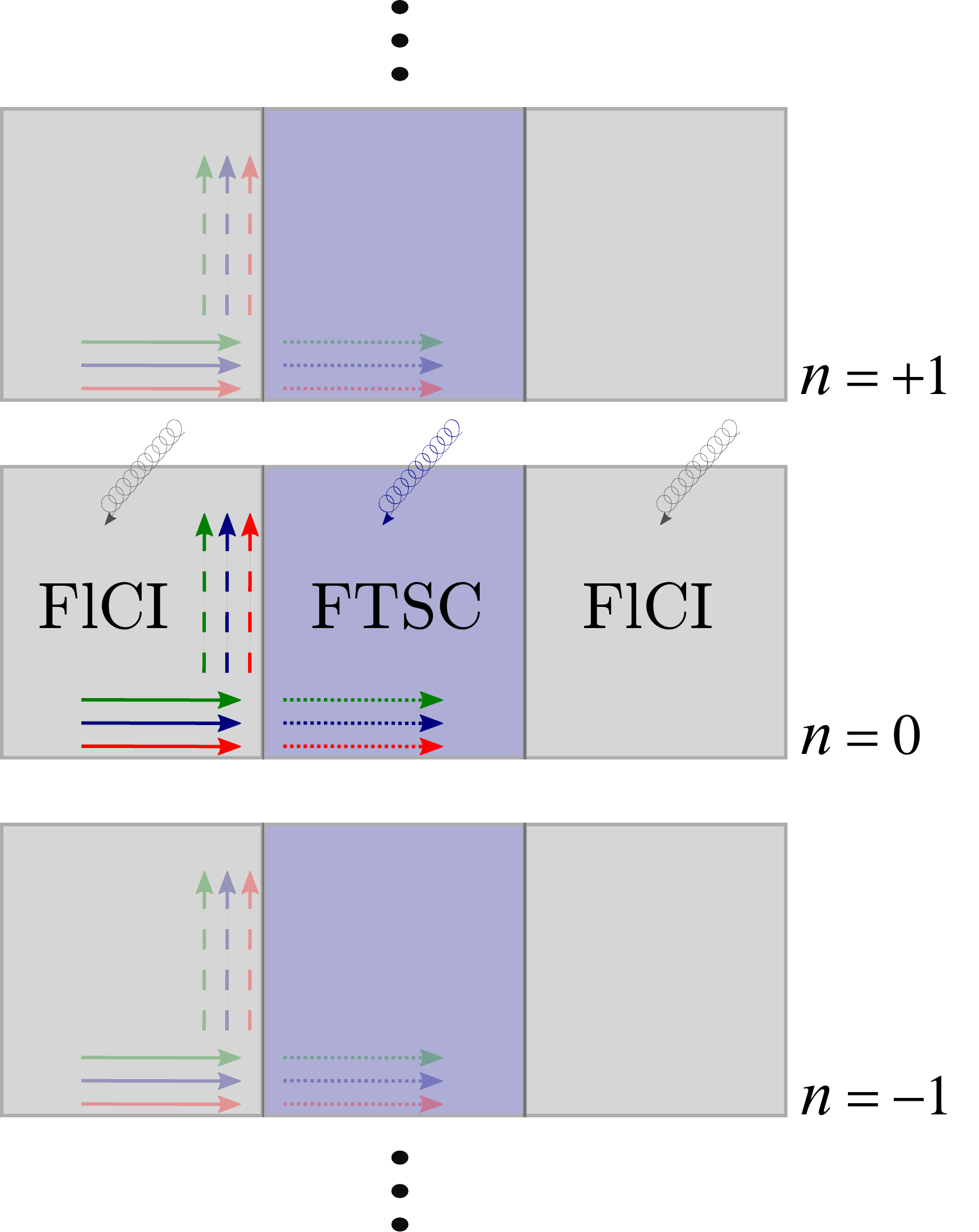}
\caption{Three incoming Floquet chiral fermions (solid arrows) split into three Floquet CMEMs along the FlCI-FTSC domain wall (dashed arrows) and three Floquet CMEMs along the FTSC-vacuum interface (dotted arrows). On top of this, the periodic drives scatter these modes into different Floquet sidebands ($n=\dots,-1,0,1,\dots$), corresponding to energies $E_\mathrm{F}+n\hbar\Omega$ at quasienergy $\epsilon_\mathrm{F}=E_\mathrm{F}$.}
\label{fig:sumrule}
\end{figure}

The picture of an incoming chiral regular fermion splitting into two CMEMs\cite{SCZhang3,SCZhang4} can then be generalized to the case of multiple edge modes. For Floquet systems, on top of this there is an additional Floquet replication in the frequency space, because the periodic driving fields scatter the modes into different Floquet sidebands\cite{Farrell}. A schematic is given in Fig.~\ref{fig:sumrule}. If transport measurements involving summations over Floquet sidebands are performed, the stringent requirement that \emph{each} sideband has equal scattering processes [Figs.~\ref{fig:graphene}--\ref{fig:Pi}(c-d)] can likely rule out alternative mechanisms\cite{XGWen,JaySau} for the observation of half-quantized DC conductance.


\subsection{Perspectives on DC transport in Floquet hybrid device}
When studying transport in heterostructures consisting of different Floquet topological matter, a critical requirement is that the device must be driven \emph{in its entirety}. We shall take the FlCI-FTSC-FlCI junction to illustrate this point. If instead of FlCIs, we have static QAHIs (or CIs) on the two sides, then either (i) they serve as leads, or (ii) they are parts of the sample that connect to metallic leads. For case (i), since the QAHIs only have a finite gap, the portions of DC conductance scattered to $n\neq 0$ Floquet sidebands cannot be salvaged by the Floquet sum rule, unless the gap is much larger than the drive frequency\cite{HHYap}. However, this would imply a large hopping mismatch between the QAHI leads and the FTSC sample which inhibits transport.
Consider now case (ii). If the QAHIs are part of the sample, which contains a driven TSC, then Floquet theory demands that the QAHIs be treated as being trivially driven, i.e. they form identical copies in the Floquet space that do not couple with each other. Nevertheless, the frequency term is nonzero (because the TSC is driven), so there is an infinitude of possibilities (corresponding to different frequencies) for the Floquet zone to be folded. Therefore, unless the frequencies are sufficiently large for the QAHIs to remain insulating after zone folding, one cannot expect the static topological features of the QAHIs to directly carry over in presence of a driven TSC. Yet large frequencies only slightly renormalize some system parameters (except for gapless systems discussed in Sec.~\ref{sec:graphene}), so that truly interesting Floquet systems such as higher plateaus and $\pi$ modes will not take place anyway.

\section{Summary}\label{sec:Summary}
{  In this work, we have investigated the two-terminal transport of Floquet chiral Majorana fermions in several model topological insulator-superconductor heterostructures.  The main finding is that upon invoking the Floquet sum rule, Floquet CMEMs not only admit the intriguing $\frac{1}{2}\frac{e^2}{h}$ conductance plateaus that are currently under extensive investigations \cite{XGWen,JaySau,SCZhangReply},  they can also yield  conductance plateaus quantized at larger half-integers.  This study thus makes a computational observation of  higher-than-$1/2$ conductance plateaus in hybrid topological junctions and demonstrates the possibility, at least in principle, of creating multiple pairs of Floquet CMEMs in FTSC.
While alternative mechanisms not based on CMEMs may also yield
conductance plateaus at $\frac{1}{2}\frac{e^2}{h}$ \cite{XGWen,JaySau},   {it is unlikely that these mechanisms can} produce
higher half-quantized plateaus upon a subtle sum rule from models with a minimal number of bands.
Other detailed results obtained in this work can also guide future quantum transport studies of Floquet topological devices. The finding that Floquet CMEMs admit equal probabilities for normal transmission and local Andreev reflection \emph{at all Floquet sidebands} may   {serve as a first test in experiments prior to observing the more challenging quantized plateaus.}}

\section*{Acknowledgement}
H.~H.~Yap thanks Nicholas Sedlmayr for helpful email exchanges.  We also thank Raditya Weda Bomantara, Linhu Li, Sen Mu, and Tianshi Xiong for discussions. J.G. is supported by the Singapore NRF grant No. NRF-NRFI2017-04 (WBS No. R-144-000-378-281) and by the Singapore Ministry of Education Academic Research Fund Tier I (WBS No. R-144-000-353-112).
\bibliographystyle{apsrev4-1}
\bibliography{FloquetMajorana}

\end{document}